\documentclass[reprint,showpacs,twocolumn,superscriptaddress,aps,nofootinbib,colorlinks=true,allcolors=blue]{revtex4-2}
\usepackage{amsmath,amsthm,amssymb}
\usepackage{graphicx}
\usepackage{physics}
\usepackage[page]{appendix} 
\usepackage[shortlabels]{enumitem}
\usepackage[most]{tcolorbox}
\usepackage{framed}
\usepackage[utf8]{inputenc}
\usepackage{listings} 
\usepackage{orcidlink}

\makeatletter
\def\section{\@startsection
  {section}{1}{\z@}%
  {-3.5ex \@plus -1ex \@minus -.2ex}%
  {2.3ex \@plus .2ex}%
  {\normalfont\bfseries\raggedright}}
\makeatother

\usepackage{tabularray}
\usepackage[american,british]{babel}
\usepackage[T1]{fontenc} 
\usepackage{bm}
\usepackage{verbatim}
\usepackage{braket}

\definecolor{metalblue}{HTML}{397378}
\definecolor{darkgreen}{HTML}{0F1E23}
\definecolor{softgreen}{HTML}{173035}
\definecolor{brightgreen}{HTML}{E1F6E9}
\definecolor{mintgreen}{HTML}{00C887}

\usepackage[most]{tcolorbox}
\usepackage[table]{xcolor}

\definecolor{okabeblue}{HTML}{0072B2}
\colorlet{okabebluebg}{okabeblue!15}

\definecolor{okabeorange}{HTML}{D55E00}
\colorlet{okabeorangebg}{okabeorange!15}

\definecolor{okabepurple}{HTML}{CC79A7}
\colorlet{okabepurplebg}{okabepurple!15}

\definecolor{okabegreen}{HTML}{009E73}
\colorlet{okabegreenbg}{okabegreen!15}

\begin{document}

\title{Noise-aware emulation and cross-device validation of neutral atom analog quantum processing units}

% --- Co-First Authors ---
\author{Constantin Dalyac\textsuperscript{$\|$}~\orcidlink{0000-0002-0339-6421}}
\email{constantin@pasqal.com}
\author{Sergi Juli\`{a}-Farr\'e\textsuperscript{$\|$}~\orcidlink{0000-0003-4034-5786}}
\email{sergi.julia-farre@pasqal.com}
\author{Lucas Leclerc\textsuperscript{$\|$}~\orcidlink{0000-0003-0581-9165}}
\email{lucas.leclerc@pasqal.com}
\author{Vittorio Vitale\textsuperscript{$\|$}~\orcidlink{0000-0003-4207-9274}}
\email{vittorio.vitale@pasqal.com}
\affiliation{Pasqal, 24 rue Emile Baudot, 91120 Palaiseau, France}

% --- Core Team (Alphabetical by Last Name) ---
\author{Boris Albrecht~\orcidlink{0000-0003-0733-2676}}
\author{Lucas B\'{e}guin~\orcidlink{0000-0003-1388-0791}}
\author{Kemal Bidzhiev}
\author{Petru Borta}
\author{Cl\'{e}mence Briosne-Frejaville~\orcidlink{0009-0007-2615-5050}}
\author{Daniel J. Campbell~\orcidlink{0000-0001-8364-0086}}
\author{Dorian Claveau~\orcidlink{0009-0003-2694-1087}}
\author{Makrem Chatti}
\author{Antoine Cornillot\,\orcidlink{0009-0004-7228-541X}}
\author{Julius de Hond\,\orcidlink{0000-0003-2217-934X}}
\author{Anita Devi}
\author{Thomas Eritzpokhoff~\orcidlink{0009-0000-3899-8564}}
\author{Gaétan Hercé}
\author{Fergus Hayes~\orcidlink{0009-0000-8139-5478}}
\author{Soufiane Kaghad}
\author{Arun Kumar Abhimanyu}
\author{Lucas Lassabli\`ere~\orcidlink{0000-0001-8081-1054}}
\author{Mauro Mendizabal}
\author{Anton Quelle}
\author{Julien Ripoll~\orcidlink{0009-0004-5282-9942}}
\author{Henrique Silvério}
\author{Joseph Vovrosh~\orcidlink{0000-0002-1799-2830}}
\author{Guillaume Villaret~\orcidlink{0000-0002-3898-8646}}
\author{Adrien Signoles~\orcidlink{0000-0001-7822-9444}}
\author{Alexandre Dauphin~\orcidlink{0000-0003-4996-2561}}
\email{alexandre.dauphin@pasqal.com}
\affiliation{Pasqal, 24 rue Emile Baudot, 91120 Palaiseau, France}

\begin{abstract}
Analog quantum processors based on Rydberg atom arrays are a powerful platform for many-body quantum simulation, combinatorial optimization, and graph machine learning. As these devices become increasingly accessible, establishing confidence in their outputs requires predictive models that quantitatively connect microscopic hardware imperfections to empirical results. Here, we present a noise-aware emulation framework that propagates the dominant noise mechanisms throughout the full computation cycle to predict device behavior. We validate the framework by benchmarking two representative protocols, quantum annealing and post-quench dynamics, on three Pasqal quantum processors where classical simulations still provide ground truth. Across all three devices, the measured observables fall within the uncertainty envelopes predicted by the emulator. Beyond reproducing the data, the framework isolates which physical mechanism dominates in each operating regime, provides quantitative guidance for algorithm design and hardware improvements, and establishes a foundation for verifying analog processors in regimes beyond classical reach.
\end{abstract}

\maketitle
\begingroup
\renewcommand{\thefootnote}{$\|$}
\footnotetext{These authors contributed equally to this work.}
\endgroup

\section{Introduction}
Neutral-atom quantum processing units (QPUs) have emerged as a leading platform for many-body quantum simulation~\cite{browaeys_many-body_2020,henriet_quantum_2020,bernien_probing_2017,menssen_strategic_2026}, with further applications in combinatorial optimization and graph machine learning~\cite{dalyac_graph_2024}. They natively encode a broad range of interacting spin models, and their programmable geometries and tunable interactions give access to system sizes and interaction regimes beyond the reach of exact classical simulation~\cite{vovrosh_simulating_2026,leclerc_one--one_2026}, which is also what makes their outputs difficult to check.

As these QPUs become more accessible, they are increasingly operated in regimes at the edge of classical simulability, where an exact reference is not routinely available. The dominant noise sources are individually well characterized at the single-atom level~\cite{de_leseleuc_analysis_2018,wurtz_aquila_2023}, but their combined effect on many-body observables is much harder to anticipate. So far, this has been addressed only through specific analyses of individual experiments~\cite{shaw_benchmarking_2024,scholl_quantum_2021}, while a general, predictive framework, backed by an open-source noise emulator, is still missing.

A predictive noise model is only part of the challenge. Even a well-characterized device tells us little on its own if its results cannot be reproduced elsewhere: when several QPUs of the same generation are meant to be used interchangeably, one needs to know that an outcome reflects the Hamiltonian that was programmed, and not the specifics of one particular machine. Comparing devices is therefore a genuine test of whether the underlying physics is reproducible across hardware.

There are, nonetheless, good reasons to expect that reliable validation remains possible even without a classical reference. While the global many-body fidelity decays rapidly with system size and noise~\cite{shaw_benchmarking_2024}, the local observables that are of practical interest remain far more robust, with errors that stay bounded when increasing the system size~\cite{trivedi_quantum_2024,flannigan_propagation_2022}. Provided the noise is well controlled and combined with moderate system sizes and error mitigation, one can therefore validate a noise model on small, classically tractable instances and extrapolate its predictions to the larger regimes where no classical reference is available.

In this work, we develop a general framework for modeling and validating noise in neutral-atom quantum simulators, together with an open-source noise emulator built on Pulser~\cite{pasqal_pulser_docs,pasqal_emumps_docs}, an open-source Python library for controlling and simulating neutral-atom devices. We apply it to three quantum processors of the same generation: FC1 and SA1, accessed through the cloud in Canada and Saudi Arabia, and Ruby, installed at the TGCC facility at CEA in France and accessed through its HPC infrastructure.

Comparing the three, we find that a single noise model, calibrated on one device, reproduces the behavior of all three. This lets us separate effects that are reproducible across devices from residual, device-specific deviations, attribute the dominant discrepancies to specific noise channels, and project how performance should improve as the hardware matures.

The paper is organized as follows. Section~\ref{sec:qpu_cycle} describes the operating cycle of a neutral-atom QPU. Section~\ref{sec:cross_benchmark} introduces the two benchmark protocols — quantum annealing and post-quench dynamics — together with the observables used to characterize them. Section~\ref{sec:qpu_noise} presents the noise model and its open-source emulator. Sections~\ref{sec:annealing} and~\ref{sec:quenches}  apply the framework to the annealing and post-quench regimes, comparing emulation against FC1 data. Section~\ref{sec:conclusions} summarizes our findings and discusses their implications for scaling neutral-atom simulators.

\begin{figure*}[t]
    \centering
\includegraphics[width=\linewidth]{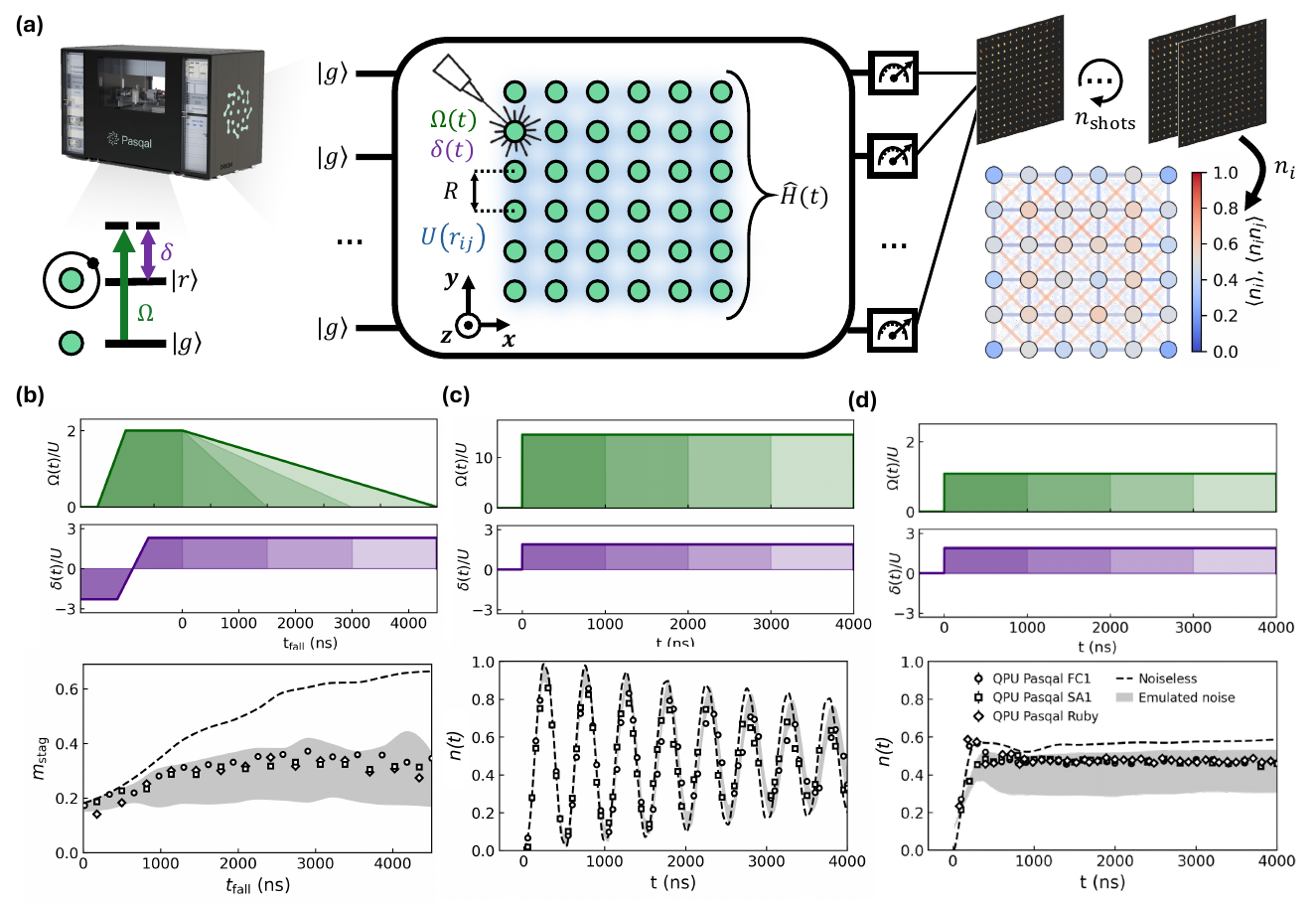}
    \caption{\textbf{Analog Rydberg QPU cycle and benchmark protocols.}
    \textbf{(a)}, Schematic of the QPU cycle for a register of two-level atoms with ground state $\ket{g}$ and Rydberg state $\ket{r}$ driven by a global Rabi frequency $\Omega(t)$ and detuning $\delta(t)$, with interactions $U_{ij}$ set by the geometry and atomic spacing $R$. After the programmed evolution, projective readout in the $z$ basis is obtained from fluorescence imaging. Bitstrings over $n_{\mathrm{shots}}$ repetitions can be used to estimate site-resolved observables such as local occupations $\langle n_i\rangle$ and correlators.
    \textbf{(b)}, Annealing protocol. Top: time-dependent control schedule with $\Omega(t)$ and $\delta(t)$. Bottom: staggered magnetization $m_{\mathrm{stag}}$ measured at the end of the evolution as a function of the final ramp-down duration $t_{\mathrm{fall}}$ of $\Omega(t)$. Experimental results from the Pasqal QPUs---FC1, SA1 and Ruby (circle, square and diamond markers, respectively)---are compared with numerical simulations. Dashed lines denote ideal (noiseless) predictions, while shaded bands indicate the uncertainty within the full noise model from Pasqal FC1. 
    \textbf{(c)}, Post-quench dynamics in the weakly interacting regime. Top: control schedule with constant $\Omega$ and $\delta$ (with $\Omega\gg U$) after a fast ramp-up. Bottom: average Rydberg population $\langle n\rangle$ as a function of the evolution time $t$, with the same experimental and numerical conventions as in \textbf{(b)}.
    \textbf{(d)}, Same as \textbf{(c)}, but for the competing interacting regime ($\Omega\simeq U$).}
    \label{fig:qpu_cycle}
\end{figure*}

\section{Anatomy of the QPU Cycle}\label{sec:qpu_cycle}
We first describe the QPU cycle, sketched in Fig.~\ref{fig:qpu_cycle}(a), from the preparation of the atomic register to its coherent evolution and final measurement. This cycle also provides the structure for identifying where each noise mechanism enters the computation that will be described in Sec.~\ref{app:ising}.

Individual atoms are trapped in optical tweezers and arranged in programmable two-dimensional arrays, forming a register of $N$ effective two-level systems defined by the ground state $\ket{g}$ and the Rydberg state  $\ket{r}$ [Fig.~\ref{fig:qpu_cycle}(a)]. All atoms are initialized in $\ket{g}$ by optical pumping. The optical tweezers are then switched off during the quantum evolution because the Rydberg state is anti-trapped by the trapping potential and would therefore be expelled from the register. The transition between $\ket{g}$ and $\ket{r}$ is driven through an effective two-level coupling characterized by a time-dependent Rabi frequency $\Omega(t)$ and detuning $\delta(t)$ (see App.~\ref{app:2level}). The resulting many-body dynamics is described by the Rydberg Hamiltonian
\begin{equation}\label{eq:rydberg_ham}
\hat{H}(t) = \sum_i \left[\frac{\hbar\Omega(t)}{2}\hat{\sigma}_i^x - \hbar\delta(t) \hat{n}_i \right]
+ \sum_{i<j} U_{ij}\, \hat{n}_i \hat{n}_j,
\end{equation}
where $\hbar$ is the reduced Planck's constant, $\hat{n}_i = (1+\hat{\sigma}_i^z)/2$ projects onto the Rydberg state and $U_{ij} = C_6/r_{ij}^6$ describes the van der Waals interaction between atoms separated by a distance $r_{ij}\equiv |\mathbf{r}_i-\mathbf{r_j}|$, where $\mathbf{r}_i$ the position vector of the $i$-th atom in the register. We define the register spacing $R$ as the minimum interatomic distance between two atoms and the interaction energy scale as $U\equiv C_6/R^6$. The geometry of the register, encoded in $\{\mathbf{r}_i\}$, directly determines the interactions between atoms, which is static, whereas the terms $\Omega(t)$ and $\delta(t)$ can be modulated during the sequence. In general, the flexibility of the platform allows one to explore all the physically relevant regimes of the Hamiltonian in Eq.~\eqref{eq:rydberg_ham}, as described in App.~\ref{app:parameters_ranges}, where we also provide a summary on the well-known connection with the quantum Ising model. Regarding time-scales, pulse sequences are typically limited to $6\,\mu$s, and the temporal profile of $\Omega(t)$ and $\delta(t)$ can be tuned either with an acousto-optic modulator (fastest ramp up of $100-200$ ns) or an electro-optic modulator (fastest ramp up of $10$ ns).

After a programmed time evolution, the traps are turned on again and atoms in the $\ket{r}$ state are expelled from the register. Subsequently, one detects the atoms in the $\ket{g}$ state by fluorescence imaging, leading to a projective quantum measurement of the local population. Repeating the experiment for $n_{\mathrm{shots}}$ independent realizations produces a set of bitstrings from which $z$-basis observables can be estimated statistically. 
In particular, besides the local populations $\langle n_i \rangle$, one can also compute $k$-body correlations $\langle n_i n_j \dots n_k\rangle$. Crucially, the QPU provides experimental access to the bitstring statistics in the large-$N$ regime, where classical numerical methods are typically unable to faithfully reproduce the corresponding many-body probability distribution because of the exponential growth of the Hilbert space. While finite sampling and hardware noise limit the characterization of the full distribution, the experimentally accessible statistics are nevertheless expected to provide more reliable information about many physically relevant observables than classical simulations at sufficiently large system sizes, thereby enabling applications in quantum simulation, machine learning, and optimization.

\section{Canonical protocols on multiple PASQAL QPUs}\label{sec:cross_benchmark}
To assess reproducibility across QPUs of the same generation, we implement two benchmark protocols on three Pasqal devices: FC1 and SA1, accessed through the cloud in Canada and Saudi Arabia, respectively, and Ruby, installed at CEA’s TGCC in France and accessed through its HPC infrastructure. We consider two prototypical protocols that serve as building blocks for applications in quantum simulation, machine learning, and optimization: quantum annealing for adiabatic state preparation and post-quench dynamics. Both protocols are implemented with $36$ atoms in  the ground state $\ket{\psi_0}=\ket{g}^{\otimes N}$ arranged on a $6\times6$ square lattice, and the relevant observables are estimated from repeated measurements at each experimental setpoint. In the numerical emulations, we use a single set of noise parameters calibrated on FC1. This set is used for comparison for all three devices without device-specific refitting, as the noise characteristics are expected to be similar across the devices. This provides a direct test of whether the model transfers across hardware instances; the parameters are summarized in Table~\ref{tab:noise_sources_hamiltonian} and detailed in App.~\ref{app:noise}.

\subsection{Quantum annealing protocol: antiferromagnetic state preparation}\label{ssec:annealing-protocol}

Quantum annealing is a natural application of analog neutral-atom QPUs: rather than decomposing the evolution into digital gates, it uses the continuous-time dynamics generated by the native many-body Hamiltonian to prepare correlated low-energy states. This approach is relevant to quantum simulation, including the preparation of phases of interest~\cite{ebadi_quantum_2021,king_beyond-classical_2025,leclerc_one--one_2026} and states near critical points~\cite{fang_probing_2025,sun_experimental_2026}, as well as to optimization, where low-energy configurations of an effective Ising Hamiltonian encode candidate solutions to combinatorial problems~\cite{ebadi_quantum_2022,leclerc_implementing_2025,bapst_quantum_2013}. The protocol is guided by the adiabatic theorem: a system initialized in an eigenstate of the Hamiltonian remains close to the corresponding instantaneous eigenstate when the Hamiltonian varies sufficiently slowly compared with the relevant inverse-gap scale~\cite{amin_consistency_2009}.

We consider as a benchmarking task the antiferromagnetic state preparation, whose protocol follows the annealing schedule from Ref.~\cite{scholl_quantum_2021} [Fig.~\ref{fig:qpu_cycle}(b) (top)], and is implemented through a piecewise-linear modulation of the global controls $\Omega(t)$ and $\delta(t)$. Concretely, $\Omega(t)$ is first ramped up from $0$ to $\Omega_{\max}=2U$ over $1000\,\mathrm{ns}$ while $\delta(t)$ is kept constant at a large negative value $\delta_i=-3U$, so that $\ket{\psi_0}$ is the ground state of the instantaneous Hamiltonian $\hat H(t=0)$; during the subsequent stage of duration $1000\,\mathrm{ns}$, $\Omega(t)$ is held constant while the detuning is swept linearly from $\delta_i$ to a positive endpoint $\delta_f$ chosen as $\hbar\delta_f=1/2\sum_{i\neq j}U_{ij}$. Finally, $\delta(t)$ is kept fixed at $\delta_f$ while $\Omega(t)$ is ramped down to zero over a variable duration $t_{\mathrm{fall}}$, which is used as the main control parameter to tune the adiabaticity of the final approach to the classical Ising limit. In terms of the effective transverse-field Ising description, this schedule connects an initial drive-dominated (paramagnetic) regime to a final interaction-dominated regime where the low-energy manifold corresponds to N\'eel order on the bipartite square lattice.

To quantify the degree of antiferromagnetic order reached at the end of the sequence, we measure the staggered magnetization 
\begin{equation}
    m_{\mathrm{stag}}
    \equiv \langle|\sum_{i\in A}\hat n_i-\sum_{i\in B}\hat n_i|\rangle/(N/2), 
    \label{eq:mstag_maintext}
\end{equation}
where $A$ and $B$ denote the two possible checkerboard sublattices \cite{scholl_quantum_2021}. This observable is directly accessible from single-site populations extracted from $n_{\rm shots}=300$ bitstring measurements, and it provides a simple diagnostic of N\'eel ordering (with $m_{\mathrm{stag}}=1$ for perfect N\'eel order in the convention adopted here). 

Figure~\ref{fig:qpu_cycle}(b) (bottom) illustrates how the antiferromagnetic order depends on the duration $t_{\mathrm{fall}}$ of the final ramp-down segment of the annealing schedule. The data extracted from the three QPUs are consistent between each other and are within the error bars of the noise model (shaded region). Increasing $t_{\mathrm{fall}}$ enhances the adiabaticity of the turn-off of the transverse drive and generally leads to a more ordered final state, as reflected by the increase of the staggered magnetization $m_{\mathrm{stag}}$. In the experimental data, this trend can be complemented by a slower saturation -- and in some cases a decrease -- at the largest $t_{\mathrm{fall}}$, consistent with the fact that longer protocols become progressively more exposed to decoherence mechanisms and other hardware imperfections. In practice, this highlights a trade-off between diabatic errors at short durations and noise-induced degradation at long durations; the noise model detailed in Sec.~\ref{sec:qpu_noise} provides a quantitative tool to identify an operating window (or ``best compromise'') and to anticipate how modifications of the schedule affect the final ordering as described in Sec.\ref{sec:annealing}.

\subsection{Post-quench protocol: quantum Ising dynamics}\label{sec:quench_protocol}
The many-body dynamics following a sudden quench of the Hamiltonian parameters represent another paradigmatic protocol in neutral atoms platforms, radically different from annealing. In this scenario, the initial QPU state has in general a finite overlap with an extensive number of eigenstates of the post-quench Hamiltonian, leading to a fast build up of quantum correlations (i.e., entanglement) away from the trivial regimes $\Omega=0$ (no dynamics) or $\Omega \gg \delta,\,U$ (single-atom Rabi oscillations).

Figure~\ref{fig:qpu_cycle}(c) (top) depicts the experimental protocol for the post-quench dynamics. We consider a constant QPU Hamiltonian, achieved by a fast ramp-up with the electro optical modulator, given by
\begin{equation}
\begin{aligned}
    &\Omega(t)=\Omega_q,\\
    &\delta(t)=\delta_q.
\end{aligned}
\end{equation}
 By choosing an atomic spacing and detuning such that $N\delta_q=\frac{1}{2}\sum_{i,j (i\neq j)} U_{ij}$, we approximately map the QPU Hamiltonian to a transverse field Ising model (see App.~\ref{app:ising} for details). In such a protocol, one is typically interested in the population dynamics
\begin{equation}
     n(t) \equiv \frac{1}{N}\sum_i \langle \hat{n}_i(t)\rangle,
 \end{equation}
which requires running the same protocol up to different final times $t$. We estimate the average over $n_{\rm shots}=200$ bitstring measurements.

Figure~\ref{fig:qpu_cycle}(c) shows the dynamics in the regime of small interactions ($\Omega/U\approx 14.5$). All devices present nearly independent single-particle Rabi oscillations, consisting on the periodic transfer in time of the population between the ground and Rydberg state. 
We also notice a slow but clearly visible decay of the oscillation amplitude in time, which can be attributed to the coherent dephasing of the many-body system due to small but finite interactions and detuning, and to noise effects which ultimately limit the coherence of the protocol at long times. Besides the amplitude behavior, we remark small offsets in the temporal periods across different devices, consistent with small residual offsets between the experimentally implemented dynamics and the programmed Hamiltonian parameters. 

Figure~\ref{fig:qpu_cycle}(d)(bottom) shows a different regime of competing driving and interactions ($\Omega/U\approx 1.1$). In contrast to the weakly-interacting case, here the single-particle picture breaks down. Instead of persistent Rabi oscillations, the Rydberg population rises rapidly at short times and then plateaus.
The experimental data show a lower plateau value with respect to the noiseless case in agreement with the emulated-noise model (gray band).

\section{Noise sources in the QPU cycle and their emulation}\label{sec:qpu_noise}
Having validated the full noise model against experimental results from three different QPUs in the previous section, we now turn to a high-level review of the different noise sources impacting a QPU cycle governed by Eq.~\eqref{eq:rydberg_ham}, together with their numerical simulation implemented in the open source library \texttt{Pulser}~\cite{silverio_pulser_2022} with backends \texttt{emu-sv} (state-vector) and \texttt{emu-mps}~\cite{bidzhiev_efficient_2025}. We therefore follow similar conventions as in the \texttt{Pulser} python library~\cite{pasqal_pulser_docs}, which takes into account previous noise descriptions in the neutral-atom literature~\cite{de_leseleuc_analysis_2018,scholl_quantum_2021-1,shaw_benchmarking_2024}.  
A more refined account of all the noise sources presented below, highlighting their relation with hardware parameters and typical values on current neutral-atom QPUs, can be found in App.~\ref{app:noise} and  Table~\ref{tab:noise_sources_hamiltonian}.

\subsection{Thermal motion of the atoms, Doppler effect, and positional disorder}\label{sec:thermal_noise}

In the register, the positional degrees of freedom of the trapped atoms can be considered to be in a low temperature $T$ thermal state, with a finite position and velocity spread both in the register plane, $xy$, and out-of-plane, along $z$. The position thermal spread leads to stochastic register positions 
\begin{equation}\label{eq:position_noise}
    \tilde{r}^\mu_i = r^\mu_i+ \mathcal{N}(\Delta r^\mu_i,\sigma^T_{r^{\mu}_i}), 
\end{equation} 
where $\mathcal{N}(\Delta r^\mu_i,\sigma^T_{r^{\mu}_i})$ denotes a random variable sampled from a normal distribution and characterizes the noise on the $\mu$-coordinate of the $i$-th register site ($\mu =x,\,y,\,z)$, with shot-to-shot thermal-induced standard deviation $\sigma^T_{r^\mu_i}$ and static mean $\Delta r^\mu_i$. 
Here, the static mean $\Delta r^\mu_i$ accounts for small systematic deviations of the programmed register positions due to imperfect laser traps, hence it is not of thermal origin.

On the one hand, the stochastic uncertainty of the position vector $\tilde{\mathbf{r}}_i$ propagates according to the Van der Waals Hamiltonian term in Eq.~\eqref{eq:rydberg_ham}, leading to stochastic interactions $\tilde{U}_{ij}$. 
On the other hand, the thermal velocity spread produces Doppler shifts and thus atom-dependent shot-to-shot fluctuations of the detuning with spread $\sigma^T_{\delta_i}\propto \sqrt{T}$. In this treatment, we neglect the position drift induced by  thermal velocities during the sequence, which is typically below $0.1\mu\textrm{m}/\mu\textrm{s}$. Further details, including the explicit temperature dependence of $\sigma^T_{r^\mu_i}$ and $\sigma^T_{\delta_i}$, and their relation with atomic and trap parameters, are provided in App.~\ref{app:thermalnoise}.

\subsection{Laser fluctuations}\label{sec:laser_noise}
The effective control parameters $\Omega(t)$ and $\delta(t)$ are derived via an off-resonant two-photon excitation scheme, as described in App.~\ref{app:2level}. Despite the level of tunability achieved in modern lasers, these are not perfectly stable and exhibit both intensity and phase fluctuations \cite{stephan_laser_2005,domenico_simple_2010}, together with spatial inhomogeneities. All of these can be considered noise sources for the effective single-photon parameters $\Omega(t)$ and $\delta(t)$. These errors can be expressed in a compact form as follows
\begin{equation}\label{eq:laser_noise}
    \begin{split}
        \tilde{\Omega}(t,r_i)&=f_\Omega(r_i)\left[\Omega(t)+\mathcal{N}(\Delta\Omega,\sigma_\Omega)\right],\\
        \tilde{\delta}(t,r_i)&=\left[\delta(t)+\mathcal{N}(\Delta\delta, \sigma_\delta)+\delta_\text{HF}(t)\right]+\mathcal{N}(0,\sigma^T_{\delta_i}).
    \end{split}
\end{equation}
Both the Rabi frequency and detuning have a stochastic contribution with a shot-to-shot standard deviation $\sigma_\Omega$ or $\sigma_\delta$, respectively. Note that the detuning spread also includes the site-dependent Doppler contribution $\mathcal{N}(0,\sigma^T_{\delta_i})$. This stochastic term can also have a non-zero mean due to small systematic biases $\Delta \Omega (\Omega, \delta)$ and $\Delta \delta (\Omega, \delta)$ away from the user-defined setpoint, and the amplitude of these biases can depend slightly on the setpoint itself. Moreover, the laser beam intensity decays away from the center of the register, leading to a slowly varying spatial envelope $f_\Omega(r_i)$. In this treatment, we neglect the spatial profile of the detuning, which comes from the dependence of this quantity with laser intensity (light shift) explained in App.~\ref{app:2level}.

Finally, laser phase fluctuations lead to a stochastic, time-dependent detuning component during the pulse sequence, $\delta_\text{HF}(t)$. During the sequence, one typically neglects time-dependent Rabi frequency fluctuations. The reason is that such fluctuations are related to intensity noise, whose power spectral density is relatively small in the relevant MHz regime, in contrast to phase noise, which is particularly enhanced.
Explicit expressions for the variables appearing in Eq.~\eqref{eq:laser_noise}, in terms of laser noise and parameters, can be found in App.~\ref{app:controls}.

\subsection{State preparation and measurement errors}\label{sec:spam}
Even if successfully trapped, an atom may not be initialized in the right computational state $\ket{g}$. The preparation error $\eta$ quantifies how many sites $i$ are erroneously prepared and thus excluded from the quantum dynamics of Eq.~\eqref{eq:rydberg_ham}, but still detected as $\ket{g}$ in the final bitstring measurement. 

At the measurement stage, readout fidelity is governed by two asymmetric processes:
the probability $\epsilon$ of misidentifying a ground state $\ket{g}$ as a Rydberg state $\ket{r}$ (e.g., due to loss of $\ket{g}$ atoms prior to imaging), and the probability $\epsilon'$ of misidentifying a Rydberg state $\ket{r}$ as a ground state $\ket{g}$ (e.g., due to decay to a ground-state before imaging). Unlike $\eta$, which directly alters the many-body dynamics, $\epsilon$ and $\epsilon'$ act as classical readout noise that can be straightforwardly mitigated when evaluating $z$-basis observables (see App.~\ref{app:SPAM}).

\subsection{Coupling to environment: effective Lindblad channels}\label{sec:lindblad}
Like any many-body quantum system, atoms are also inherently coupled to environmental degrees of freedom beyond the local two-level system. This coupling leads to decoherence of the many-body wavefunction, whose dynamics can be effectively described by a master equation \emph{\`a la Lindblad}
\begin{equation}\label{eq:Lindblad}
\dot \rho = -i[\hat H, \rho] + \sum_k \gamma_k \left( \hat L_k \rho \hat L_k^\dag - \frac{1}{2} \{ \hat L_k^\dag \hat L_k, \rho \} \right),
\end{equation}
where $[\cdot,\cdot]$ and $\{\cdot,\cdot\}$ denote the commutator and anti-commutator, respectively and $\rho$ is the density matrix.
Here, $\hat L_k$ are the jump operators associated with each dissipative channel and $\gamma_k$ are the corresponding rates. There are two main channels to be considered. The first is effective spontaneous decay from $\ket{r}$ to $\ket{g}$, characterized by the single-atom jump operator $\hat{L}_k=\ket{g}\bra{r}$ and associated decay rate $\gamma_1=1/T_1$, where $T_1$ is the effective Rydberg state lifetime (accounting for both spontaneous emission and black-body radiation). The second is dephasing in the $\{\ket{g}, \ket{r}\}$ manifold, due to adiabatic elimination or scattering from the intermediate level used to couple $\ket{g}$ and $\ket{r}$ in a two-photon scheme. This channel can be approximated by $\hat{L}_k=\ket{r}\bra{r}$ and an effective dephasing rate $\gamma_2$. Although they are beyond the scope of this work, more refined effective descriptions can include additional leakage states. 
In App.~\ref{app:channels} we provide details on the derivation of these channels.

\subsection{Review of Monte Carlo wavefunction sampling for noise emulation}\label{sec:qpu_emulation}
The impact of the noise sources described in the previous sections can be integrated in classical numerical methods. The approach we employ is described below and is implemented in the open-source \texttt{Pulser}~\cite{silverio_pulser_2022} backends \texttt{emu-sv} (state-vector) and \texttt{emu-mps}~\cite{bidzhiev_efficient_2025} (based on matrix-product-states), optimized for neutral-atom QPU emulations. In the next sections we use specifically \texttt{emu-mps}.

The main ingredient is a numerical solver of the Schrodinger equation that provides access to the QPU wavefunction and its observables and bitstring distribution at a time $t$ \begin{equation}\label{eq:sch_eq}
\ket{\psi(t)}=\mathcal{T}\left[\exp(-\frac{i}{\hbar}\int_{s=0}^{t}\hat{H}(s)ds)\right]\ket{\psi_0},
\end{equation}
where $\mathcal{T}$ is the time-ordering operator, and $\ket{\psi_0}$ is the initial QPU state, i.e., $\ket{\psi_0}=\ket{g}^{\otimes N}$ in the absence of preparation errors. With this solver, which can be readily used to emulate the noiseless QPU cycle, one can easily emulate the thermal and laser noise sources, outlined in Sections \ref{sec:thermal_noise} and \ref{sec:laser_noise} via the Monte Carlo wavefunction sampling method. In this scheme, $n_\text{MC}$ trajectories are solved according to Eq.~\eqref{eq:sch_eq}, independently sampling the distributions of the different noise sources. The trajectories final-state ensemble is then used to sample bitstrings or compute physical observables, mimicking the stochastic effect of noise in a QPU cycle. The errors described in Section \ref{sec:spam} can be straightforwardly incorporated as well:  preparation and leakage errors are taken into account by excluding specific sites from the time evolution, and associating them with the right bitstring final value. Measurement errors can be simply accounted by stochastically flipping the  bitstrings obtained from the solver. Finally, despite being beyond the two-level manifold, the effective Lindblad channels of Section \ref{sec:lindblad} can be naturally incorporated in this trajectory scheme via the quantum jump formalism.  

As anticipated above, the ensemble of $n_\textrm{MC}$ trajectories, $\{ \ket{\psi_k(t)}\}$, can be therefore used to sample stochastic QPU bitstrings~\cite{ferris_perfect_2012}, reproducing both the physical noise during the evolution, as well as noise associated to the measurement (false detections and finite-sampling). Alternatively, when neglecting measurement noise, the trajectories can also be used as direct estimators of any QPU observable $\langle \hat{O} \rangle$ as
\begin{equation}\label{eq:MC_observables}
\langle \hat{O} \rangle_\textrm{QPU}(t) \approx \frac{1}{n_\text{MC}} \sum_{k=1}^{n_\text{MC}} \langle \psi_k(t) | \hat{O} | \psi_k(t) \rangle.
\end{equation}
Here it is worth discussing the statistical interpretation of such an estimator. The standard error of the Monte Carlo mean is not, by itself, a measure of the typical variability induced by hardware noise: for independent trajectories it decreases as $n_\textrm{MC}^{-1/2}$, and therefore only quantifies how precisely the average of the chosen noise model has been estimated. To assess the fluctuations that may be observed in finite-sample noisy realizations, it is more useful to characterize the spread of the trajectory distribution itself. This spread captures the sensitivity of the observable to physical noise fluctuations and remains finite in the limit of large $n_\textrm{MC}$.
It can be summarized, for instance, by reporting percentile intervals of the values
$\langle \psi_k(t)|\hat O|\psi_k(t)\rangle$
across trajectories, or by using a bootstrap procedure designed to estimate such distributional intervals. The resulting interval should therefore be interpreted as a measure of trajectory-to-trajectory variability, rather than as the statistical error of the mean. This distinction is particularly important in the presence of correlated or slowly varying noise, for example when a fluctuation that is assumed to be independent from shot to shot becomes effectively static over a given experimental run.
In this work, we consider a spread of trajectories within two standard deviations from the mean.

Finally, we note that in real hardware a measurement can only be performed after the driving fields have been ramped-down and switched off. Since this ramp-down is deterministic, it can be emulated explicitly.
However, in many situations the computational cost can be substantially reduced by reading out $\ket{\psi_k(t)}$ at intermediate times, such that single trajectories up to a maximum time $t_\textrm{max}$ can also be used as estimators at $t<t_\textrm{max}$. Although this approximation does not generally alter substantially the results, due to fast ramp down times of few tens of ns, the associated systematic modeling caveat should be kept in mind.

\section{QPU noise in annealing protocols}\label{sec:annealing}
We now analyze the annealing protocol by applying the noise model and methodology presented in Sec.~\ref{sec:qpu_noise} using \texttt{Pulser} with its \texttt{emu-mps}~\cite{bidzhiev_efficient_2025}. Here the object of interest is the ordered bitstring prepared at the end of the sequence, so we sample bitstrings from the final state of each noisy trajectory, exactly as in the QPU readout, and estimate the staggered magnetization from their statistics. Since the observable is built directly from measured bitstrings, we include measurement errors in the numerical simulations and compare against the raw experimental data, as detailed in App.~\ref{app:SPAM}. More details on the numerical emulations are provided in App.~\ref{app:numerics}.
 
\begin{figure}
    \centering
    \includegraphics[width=\columnwidth]{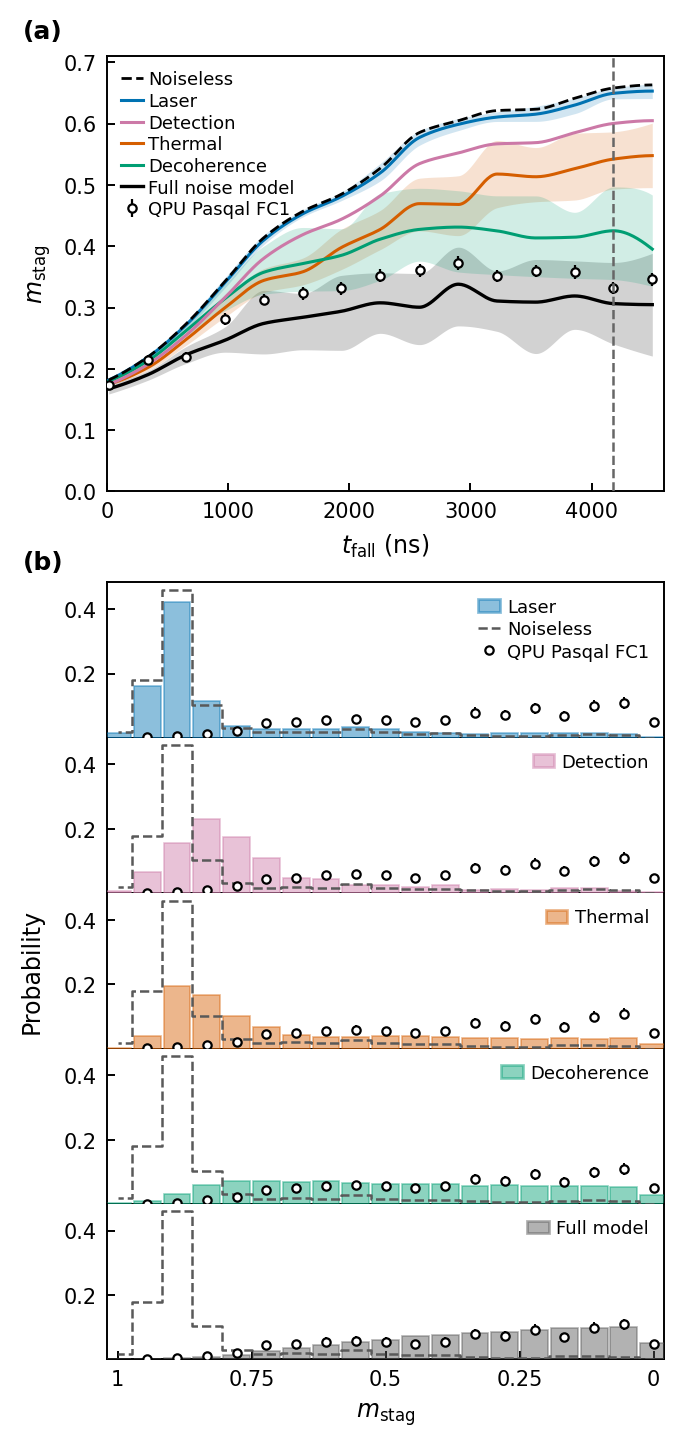}
    \caption{\textbf{
Annealing noise-budget decomposition.} \textbf{(a)}, Staggered magnetization $m_{\mathrm{stag}}$ versus final ramp-down duration $t_{\mathrm{fall}}$. Noiseless emulation (black dashed), single-noise contributions---laser (blue), detection (pink), thermal (orange), decoherence (green), ---and full noise model (black, with shaded Monte Carlo envelope), compared to FC1 experimental data (markers with shot-noise error bars). \textbf{(b)}, Outcome distribution of bitstring probability at $t_{\mathrm{fall}}=4180\,\mathrm{ns}$ ordered by decreasing $m_{\rm stag}$ for the noiseless (dashed), single-noise (colored), full noise model simulations (gray) and FC1 experimental data (markers) with error bars obtained from binning.}
    \label{fig:noise-annealing}
\end{figure}

\subsection{Impact of noise on annealing observable}

The impact of noise on annealing protocols is analyzed by applying the noise mechanisms introduced in Sec.~\ref{sec:qpu_noise} to the control schedule described in Sec.~\ref{ssec:annealing-protocol}. In contrast to the post-quench dynamics discussed later, the goal here is not to reproduce the full time dependence of the observables, but to prepare at the end of the sequence a many-body state with pronounced antiferromagnetic order, as quantified by a large staggered magnetization $m_{\mathrm{stag}}$ (with $m_{\mathrm{stag}}=1$ for an ideal N\'eel-ordered state in the convention used here).

A natural reference point is the coherent (noiseless) evolution generated by Eq.~(\ref{eq:sch_eq}). Even in this idealized limit, the final state generally does not coincide perfectly with the target ground state. The annealing path typically crosses a region where the many-body gap $\Delta(s)=|E_1(s)-E_0(s)|$ becomes small compared to the relevant dynamical scales, making strictly adiabatic evolution increasingly demanding. Since the adiabatic condition scales approximately as $t_f\propto \Delta_{\min}^{-2}$, approaching perfect adiabaticity would require evolution times that quickly exceed experimentally accessible pulse durations. The protocol therefore operates in a quasi-adiabatic regime, where Landau--Zener-type excitations are generated near the minimum gap. These diabatic excitations impose an intrinsic ceiling on the achievable staggered magnetization, already before accounting for hardware noise.

Figure~\ref{fig:qpu_cycle}(b) (bottom) reports the staggered magnetization $m_{\mathrm{stag}}$ measured after the annealing sequence on several QPUs as a function of the final ramp-down duration $t_{\mathrm{fall}}$. The three devices (FC1, SA1, Ruby) display a consistent trend: increasing $t_{\mathrm{fall}}$ enhances antiferromagnetic ordering, with $m_{\mathrm{stag}}$ rising and then saturating. This device-to-device reproducibility indicates that the protocol operates in a robust regime, where the final ordering is governed primarily by the programmed Hamiltonian path rather than by strong, hardware-specific deviations.

Importantly, the end-to-end emulation based on the full noise model reproduces the measured $m_{\mathrm{stag}}$ values across devices within experimental uncertainties. The agreement holds both for the overall growth with $t_{\mathrm{fall}}$ and for the saturation at values well below the ideal antiferromagnetic limit. This supports the interpretation that the dominant limitations can be attributed to a combination of identifiable physical noise mechanisms, as opposed to uncontrolled effects that would vary substantially from one QPU to another.

\subsection{Influence of different noise families}
The relative role of the different noise mechanisms in the annealing protocol is clarified by the decomposition shown in Fig.~\ref{fig:noise-annealing}. In panel (a), the noiseless emulation provides the coherent baseline for the achievable staggered magnetization as $t_{\mathrm{fall}}$ is increased, while the colored curves isolate the impact of adding each noise family independently. Panel (b) complements this global metric by resolving how noise redistributes probability weight among measurement outcomes: at fixed $t_{\mathrm{fall}}=4180\,\mathrm{ns}$, the bitstring probabilities show a progressive transfer of weight from highly ordered outcomes (large $m_{\mathrm{stag}}$) toward less ordered configurations (smaller $m_{\mathrm{stag}}$), providing a microscopic picture for the reduction observed in panel~(a).

The laser-noise contribution produces only a weak additional reduction of $m_{\mathrm{stag}}$. This is consistent with the general robustness of adiabatic state-preparation protocols to small variations of the Hamiltonian parameters, as well as with their relative insensitivity to the precise path followed in the phase diagram. In practice, shot-to-shot fluctuations of the programmed controls $\Omega(t)$ and $\delta(t)$ mainly act as small perturbations of the instantaneous trajectory, which only weakly alter the final antiferromagnetic order in the regime explored here. Correspondingly, panel~(b) shows that the associated redistribution of bitstring weights away from the highest-$m_{\mathrm{stag}}$ sector remains limited. Moreover, the influence of in-sequence phase noise (modeled through $\delta_{\mathrm{HF}}(t)$) is comparatively reduced in this slow-annealing setting, so that its net effect on the final ordering remains small.

Detection errors introduce an approximately systematic downward shift of the measured $m_{\mathrm{stag}}$, consistent with a partial loss of sublattice contrast at readout due to misclassification events. In panel~(b), this appears as an increased weight of intermediate-$m_{\mathrm{stag}}$ outcomes even when the underlying state remains comparatively ordered. For the calibrated error rates considered, this contribution remains subdominant compared to the dominant dynamical noise sources. In addition, when the detection-error probabilities are independently characterized, their effect on single-site and few-body observables can typically be compensated by classical postprocessing, as detailed in App.~\ref{app:SPAM}.

Thermal effects provide a substantially stronger limitation. Doppler-induced detuning fluctuations (with spread $\sigma^T_\delta$) act as site-dependent phase noise and can induce local dephasing over the course of the evolution. In parallel, thermal motion generates shot-to-shot positional disorder, which propagates to the interaction matrix $U_{ij}\propto r_{ij}^{-6}$. As a result, each realization effectively experiences a slightly distorted interaction graph, strengthening some antiferromagnetic links while weakening others; averaging over realizations then reduces the net build-up of staggered order. This mechanism is directly reflected in panel~(b) by a clear loss of probability mass from the most ordered local bitstrings toward configurations with smaller staggered magnetization.

Decoherence channels (parameterized by $\gamma_1$ and $\gamma_2$) have a pronounced impact and largely control the long-time behavior. Unlike classical parameter fluctuations, these processes generate genuine non-unitary dynamics within each realization, suppressing the coherent buildup of correlations that the annealing protocol aims to establish. In practice, this is manifested in panel~(a) by a strong reduction and saturation of $m_{\mathrm{stag}}$ at large $t_{\mathrm{fall}}$, and in panel~(b) by a substantial enhancement of low-$m_{\mathrm{stag}}$ outcomes, consistent with the proliferation of local defects and the loss of long-range order.

When all contributions are included simultaneously, the full noise model yields a quantitative prediction that closely matches the QPU measurements, as showcased in both panels of Fig.~\ref{fig:noise-annealing}, and provides the primary reference for the comparison in Fig.~\ref{fig:qpu_cycle}(b) (bottom). The finite confidence intervals indicate that the trajectories are relatively dispersed in this regime, making the observable particularly sensitive to fluctuations. Consequently, the slight discrepancies observed between the different QPUs and between the full-noise emulation and the experimental data can be naturally understood as arising from this sensitivity. Additionally, residual systematic effects not captured by the stochastic model, such as minor miscalibrations or slow drifts over time of $\Omega_{\max}$, $\delta_f$ or $R$, as well as distortions of the programmed control shapes induced by the finite bandwidth and transfer function of the AOM-based modulation may also contribute at a minor level. 

\section{QPU noise in post-quench dynamics protocols}\label{sec:quenches}
We now analyze the post-quench dynamics introduced in Sec.~\ref{sec:quench_protocol} by applying the noise model and methodology presented in Sec.~\ref{sec:qpu_noise} using \texttt{Pulser} with its \texttt{emu-mps} back-end~\cite{bidzhiev_efficient_2025}. In contrast to the annealing case, where the goal is to prepare specific bitstrings measured at the end of the sequence, the object of interest here is the full time dependence of observables during the many-body evolution. Accordingly, we make a twofold approximation in the numerical estimation of these time-dependent observables. First, we estimate observables at intermediate times using end-time trajectories, thereby neglecting the additional ramp-down time that is required to measure these in the QPU. Second, instead of sampling bitstrings, we evaluate the observables from the evolved wavefunctions following Eq.~\eqref{eq:MC_observables}.
For consistency, since measurement errors are therefore not included in the numerical simulations, we correct for their effect directly in the experimental QPU data, as detailed in App.~\ref{app:SPAM}. More details on the numerical emulations are provided in App.~\ref{app:numerics}.

\subsection{Impact of noise on post-quench population dynamics}
\begin{figure}[h]
    \centering
\includegraphics[width=1\linewidth]{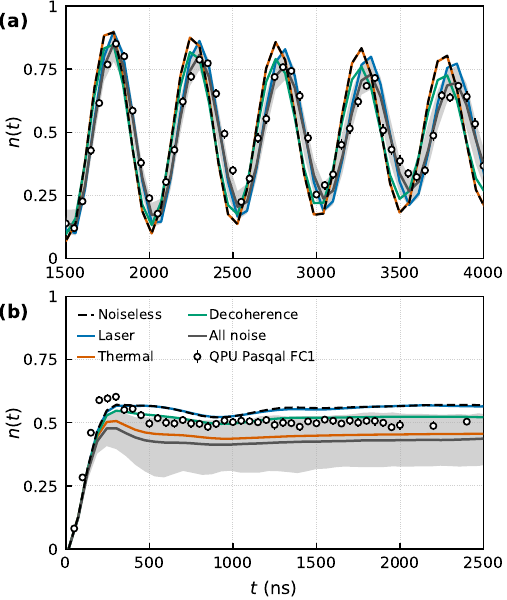}
    \caption{\textbf{Noise-emulation methodology for the post-quench dynamics protocol}. Dynamics of the mean Rydberg population, $n(t)$, in a $6\times 6$ square lattice. We compare the Pasqal FC1 data (white marker with shot-noise error bars) against \texttt{emu-mps} emulators of the ideal Hamiltonian (dashed black line), the full noise model (solid green line, with confidence area), and marginal noise models with effects from: lasers (solid blue line), temperature (solid red line), and effective channels (solid gray line). For a better visualization, we did not include the confidence area of the marginal noise models. \textbf{(a)}, Small interaction regime, $\Omega/U=14.5$. \textbf{(b)}, Strongly interacting regime, $\Omega/U=1.1$.}
    \label{fig:quench_noises}
\end{figure}
We focus first on the mean Rydberg population $n(t)$.
Local observables of this type are directly estimated from the measured QPU bitstrings with a modest shot budget and constitute the elementary building blocks of most quantum-simulation applications. They therefore provide a natural first diagnostic of how hardware imperfections affect the programmed dynamics. A natural reference point is again the coherent (noiseless) evolution generated by Eq.~\eqref{eq:sch_eq}, whose behavior differs qualitatively between the two parameter regimes of Figs.~\ref{fig:qpu_cycle}(c,d), also shown in Figs.~\ref{fig:quench_noises}(a,b). 
In the weakly interacting regime ($\Omega/U=14.5$), the system remains close to a collection of independently driven atoms, and $n(t)$ displays large-amplitude Rabi oscillations; small but finite interactions and detuning nevertheless induce a slow, purely coherent damping of the oscillation contrast.
In the regime of competing drive and interaction ($\Omega/U=1.1$), the single-atom picture breaks down and the population rises rapidly before relaxing toward a quasi-stationary plateau. Part of the damping visible in the ideal curves is thus intrinsic to the coherent many-body dynamics and must not be attributed to hardware noise; disentangling the two contributions is precisely what the noise-aware emulation enables.

Fig.~\ref{fig:qpu_cycle}(c,d) shows that the three devices reproduce the same qualitative features in both regimes: persistent oscillations with slowly decaying contrast for weak interactions, and a fast rise followed by a plateau in the competing regime. The residual device-to-device differences, such as small offsets of the oscillation period, are compatible with small systematic deviations of the implemented parameters from their programmed values. As for the annealing benchmark, this reproducibility indicates that the measured dynamics is governed primarily by the programmed Hamiltonian rather than by strong hardware-specific deviations.

Importantly, the emulation based on the full noise model reproduces the measured population dynamics within the Monte Carlo uncertainty envelopes. The agreement holds both for the reduced oscillation contrast in the weakly interacting regime and for the lowered plateau in the strongly interacting one, supporting the interpretation that the dominant deviations from the ideal dynamics can be attributed to a combination of identifiable physical noise mechanisms.

\subsection{Influence of different noise families}
 
The relative role of the different noise mechanisms is studied in Fig.~\ref{fig:quench_noises}, where the noiseless emulation (dashed line) is compared with marginal noise models in which a single noise family --- e.g. laser fluctuations, thermal effects, or effective decoherence channels --- is activated, together with the full noise model.
The central observation is that the impact of each mechanism depends strongly on the physical regime: for weak interactions (Fig.~\ref{fig:quench_noises}(a)), the dynamics remains close to an effectively single-particle Rabi regime and is comparatively robust, whereas in the competing regime (Fig.~\ref{fig:quench_noises}(b)) the dynamics is intrinsically more sensitive to fluctuations of the Hamiltonian parameters and to decoherence. This distinction is essential for interpreting the relative importance of the different noise channels.

The laser-noise contribution combines shot-to-shot fluctuations of the Rabi frequency and detuning ($\sigma_\Omega$, $\sigma_\delta$) with the in-sequence phase noise $\delta_{\mathrm{HF}}(t)$. In the weakly interacting regime, these fluctuations act as a fictitious dephasing between coherent realizations: the individual trajectories slowly desynchronize, and the ensemble average exhibits a mild reduction of the oscillation contrast with respect to the noiseless curve. A similar effect also arises from the finite waist of the excitation laser, which induces a spatial dependence of the Rabi frequency across the array. Atoms located away from the beam center therefore oscillate at slightly lower frequencies, leading to an additional inhomogeneous dephasing in the averaged signal. In the strongly interacting regime, by contrast, the effect on the mean population is barely visible: once the system has relaxed toward its quasi-stationary value, the population is insensitive to small fluctuations of the control parameters. We stress that this robustness concerns stochastic fluctuations only but systematic setpoints biases can still displace the dynamics, particularly at early times, as discussed in the next subsection.

Thermal effects display the opposite regime dependence. Doppler shifts introduce site-dependent detuning fluctuations of spread $\sigma^T_\delta$, while thermal motion generates shot-to-shot positional disorder ($\sigma^T_{r^{\mu}_i}$) that effectively modifies the interaction matrix through $U_{ij}\propto r_{ij}^{-6}$. In the weakly interacting regime this contribution is almost negligible: the interatomic distances are large enough that the interaction scale remains small compared with the Rabi frequency, and the corresponding curve is barely distinguishable from the noiseless one. In the strongly interacting regime, positional disorder becomes the dominant limitation: each realization evolves under a slightly distorted interaction graph, and averaging over realizations produces a marked shift of the population plateau. Indeed, the full-noise prediction closely follows the thermal curve, indicating that thermal disorder is the most prominent in this regime.

The decoherence channels act differently from the previous families by generating genuine non-unitary dynamics within each realization. For the mean population, their impact is most visible in the weakly interacting regime, where dephasing directly damps the coherent oscillations and contributes to the reduction of the contrast over the accessible time window. In the strongly interacting regime, instead, the population has already relaxed toward a quasi-stationary value which dephasing barely modifies.

When all contributions are included simultaneously, the full noise model yields a quantitative prediction that matches the QPU measurements over most of the explored time window. In the weakly interacting regime, the full noisy curve remains close to the noiseless prediction, with a reduced oscillation amplitude and a Monte Carlo envelope that broadens with time and encompasses the measured points. In the strongly interacting regime, the prediction departs systematically from the ideal dynamics: the population plateau is shifted downward-and the late-time QPU data indeed settle within the noisy envelope, below the noiseless curve. At early times, however, the measured population lies above the full-noise prediction, close to or even above the noiseless curve; this residual discrepancy cannot be captured by zero-mean fluctuations and points to small systematic offsets of the Hamiltonian parameters beyond calibration precision, which we analyze in the following subsection.

\subsection{Accounting for residual systematic offsets in QPU–noise model comparisons}
Up to this point, we have discussed noisy numerical emulations where the fluctuating parameters were implicitly assumed to be centered around their calibrated setpoints. We now relax this assumption and incorporate biased noise, i.e., static systematic offsets in QPU parameters that lie beyond the precision of their calibration procedure. 

\begin{figure}
    \centering
    \includegraphics[width=1\columnwidth]{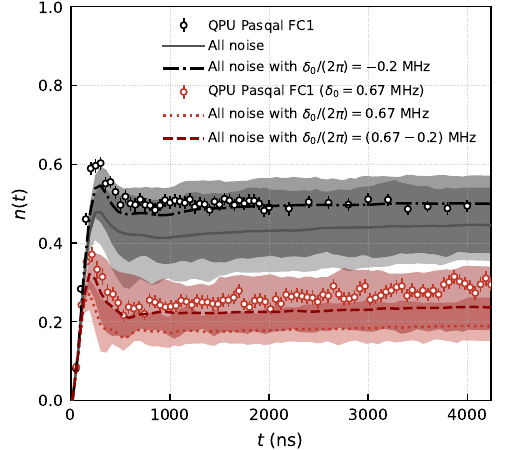}
    \caption{\textbf{Impact of systematic experimental parameter biases in post quench-dynamics}. Dynamics of the mean Rydberg population, $n(t)$, in a $6\times 6$ square lattice (strong interaction regime $\Omega/U=1.1$). Emulations  of the full noise model with \texttt{emu-mps}, including different systematic global detuning biases, $\delta_0=-0.2, 0.67, 0.47$ MHz, are shown with confidence intervals. Two Pasqal FC1 data are shown for comparison: one at the target setpoint and one where an offset $\delta_0=0.67$ MHz was intentionally programmed (orange dots).}
    \label{fig:offsets}
\end{figure}

Figure~\ref{fig:offsets} illustrates how, in certain sensitive regimes, such offsets might be needed in the noise model to remove residual discrepancies with the QPU data. We focus on the impact of detuning offsets $\Delta\delta$ in the case $\Omega/U=1.1$, which represents an intermediate regime where interactions and the Rabi frequency compete. We observe that the early-time discrepancy between the noise model and the QPU data is compatible with the presence of a $\Delta\delta/(2\pi)=-0.2$ MHz offset. It is worth stressing that such a systematic offset is, by definition, not known \emph{a priori}, and therefore one cannot rule out that similar systematic offsets (e.g., in $\Omega$ or $r_{ij}$) could be playing a comparable role. At the same time, as discussed in Sec.~\ref{sec:qpu_emulation}, the trajectory percentile already provides a useful indication of the sensitivity of the dynamics to such static parameter variations, since these effects are partially reflected in the ensemble of sampled trajectories. Nevertheless, focusing on a single parameter is typically already sufficient to quantify the dependence of a given observable on Hamiltonian offsets, while avoiding the considerably more involved task of sampling unknown static offsets on top of shot-to-shot fluctuations.

Since the value $\delta_0/(2\pi)=-0.2$~MHz is inferred a posteriori from the comparison itself, we performed an independent consistency check of this interpretation: we deliberately programmed a detuning offset, $\delta_0/(2\pi)=+0.67$~MHz, much larger than both the typical calibration precision ($\approx 100$~kHz, Tab.~\ref{tab:noise_sources_hamiltonian}) and the inferred residual bias and repeated the experiment (red circles in Fig.~\ref{fig:offsets}). This test probes two distinct aspects of the analysis. First, it verifies that the noise model correctly captures the sensitivity of the observable to detuning biases: the emulation evaluated at the programmed offset (red dotted line) quantitatively predicts the strong suppression of the population plateau. Second, it corroborates the residual-bias hypothesis itself: the measured population again lies slightly above the corresponding prediction.  A single residual detuning bias thus explains both datasets simultaneously.

\subsection{Noise effects on quantum correlations}
\begin{figure}
    \centering
    \includegraphics[width=1\linewidth]{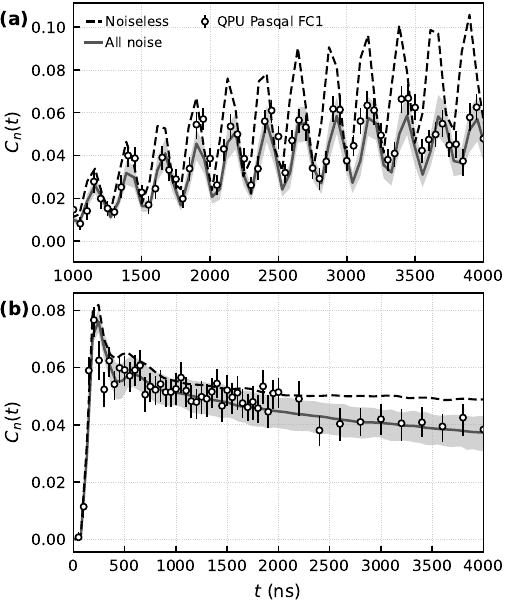}
    \caption{\textbf{Impact of noise in post-quench dynamics correlations.} Dynamics of the mean nearest-neighbor correlation $C_n(t)$. We compare the Pasqal FC1 data with \texttt{emu-mps} emulation of the ideal Hamiltonian (dashed line) and the full noise model (solid line). \textbf{(a)}, Weak interaction regime. \textbf{(b)}, Strong interaction regime.}
    \label{fig:noise_quench_connected}
\end{figure}

Let us now consider the impact of QPU noise on the growth of quantum correlations during the post-quench dynamics. To this aim, in Fig.~\ref{fig:noise_quench_connected} we show the dynamical evolution of the nearest-neighbor correlation in the lattice, $C_n=\frac{1}{N_{\langle ij\rangle}}\sum_{\langle i j\rangle}\langle \hat{n}_i\hat{n}_j\rangle-\langle \hat{n}_i\rangle\langle\hat{n}_j\rangle$, with $N_{\langle ij\rangle}$ being the number of nearest-neighbors. 
For small interactions (Fig.~\ref{fig:noise_quench_connected}(a)) the QPU captures a monotonic increase of this observable in time. In contrast, stronger interactions (Fig.~\ref{fig:noise_quench_connected}(b)) lead to a faster propagation followed by a slow decay. The numerical emulation allows us to certify that the QPU captures the qualitative behavior expected from the ideal Hamiltonian, and that the lower correlations observed in the QPU are consistent with the noise model prediction.  
The latter highlights the ability of the noise model to capture dynamical correlations, which are in general important since several many-body effects are not captured by single-qubit observables. At the same time, the strongly interacting regime provides a stringent test not only for the QPU and the noise model, but also for the classical emulation itself. Indeed, a bond-dimension scaling analysis (see App.~\ref{app:numerics}) indicates that the MPS results in Fig.~\ref{fig:noise_quench_connected}(b) are not fully converged already on time scales of $\approx 1000\,\textrm{ns}$. Nevertheless, these simulations remain valuable for asessing qualitative trends in the effect of noise, in a regime close to the edge of classical tractability.

\subsection{Boosting measurement sensitivity by reducing hardware noise}
\begin{figure}[h]
    \centering
    \includegraphics[width=1\columnwidth]{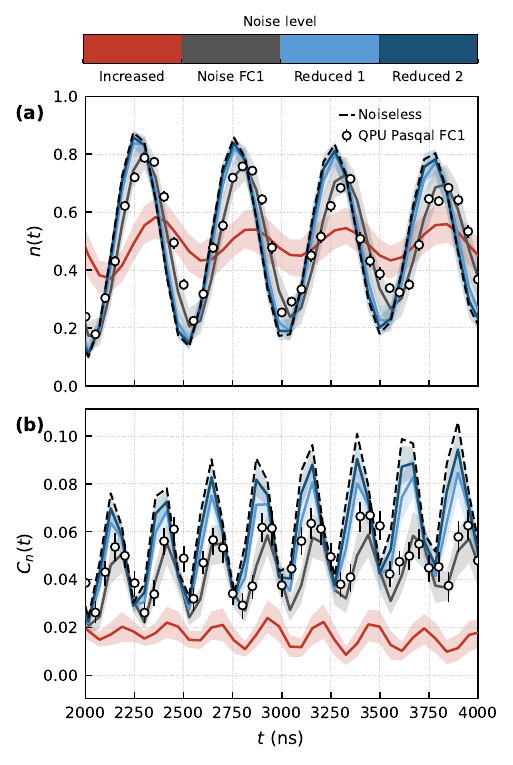}
    \caption{\textbf{Effect of noise reduction in post-quench dynamics.} Late-time dynamics of the mean occupation \textbf{(a)} and nearest-neighbor correlation \textbf{(b)} in the small interaction regime $\Omega/U=14.5$. The Pasqal FC1 data (white dots) is compared with \texttt{emu-mps} emulations of the ideal Hamiltonian (dashed black line), and the noise model with four sets of parameters (solid lines, colorcode depicted in the upper colobar): The estimated values at the time of the QPU run, and increased and reduced values (see App.~\ref{app:parameters_ranges} for details).}
    \label{fig:improvement}
\end{figure}

In addition to reproducing current device performance, the same emulation framework can be used to explore prospects under improved hardware conditions. We illustrate this procedure in Fig.~\ref{fig:improvement}(a,b). We compare the current noise emulation, matching the QPU Pasqal FC1 data, with predictions using higher (red) or lower (blue) noise levels (see App.~\ref{app:parameters_ranges} for details on the noise model parameters choices). As expected, reducing (increasing) noise levels brings the noise-aware emulation of observables closer (further) to their noiseless prediction. Another interesting point is that, while close-to-noiseless behavior is observed for the reduced noises in the local population $n$, the correlation observable $C_n$ exhibits a larger relative discrepancy, which reflects how sensitivity thresholds might depend on specific observables.  

Finally, note that these noise levels reflect worsening or improvements on the low-level hardware controls, such as laser stability or trap temperature. Improving these controls is not expected to require resources that scale exponentially with the array size. Moreover, for a fixed strength of a local noise source, its effect on local and quasi-local physical observables is expected to be controlled primarily by the evolution time, rather by the total system size~\cite{trivedi_quantum_2024}.  Together, these considerations offer a promising route towards simulating large qubit arrays with increasingly high precision, in contrast to classical numerical methods, for which the required computational resources generally grow exponentially with system size.

\section{Conclusion and Outlook}\label{sec:conclusions}
In this work, we introduced and validated a noise-aware emulation framework for analog neutral-atom QPUs. The framework propagates established microscopic noise mechanisms through the full computation cycle using Monte Carlo wavefunction sampling, as implemented in the open-source \texttt{Pulser} library and its \texttt{emu-mps} backend~\cite{silverio_pulser_2022,pasqal_emumps_docs}.

We benchmarked the framework on three Pasqal QPUs using two complementary protocols—adiabatic state preparation and post-quench dynamics—in a regime where classical simulations provide a reference. The annealing order parameter and post-quench population are reproducible across the three devices and fall within the uncertainty envelopes predicted by a single noise model calibrated on FC1, without device-specific refitting. Detailed comparisons on FC1 further show that the framework captures both local observables and connected two-body correlations. The remaining discrepancies are consistent with small systematic offsets beyond calibration precision.

Beyond validation, the framework provides a practical tool for both algorithm and hardware development. It can be used to optimize protocol parameters and assess their robustness before execution on a QPU. By decomposing the total error into distinct physical noise channels, it also identifies which hardware improvements are most likely to translate into the largest performance gains for a given application.

Looking ahead, this framework provides a route toward the reliable interpretation of analog QPU results beyond the reach of classical simulation. The strategy is to validate noise models where classical comparisons remain possible, test their transfer across devices, and then use them to assess larger systems for which exact benchmarks are unavailable.

\section*{Acknowledgments}

Pasqal acknowledges funding from the European Union under the projects PASQuanS2.1 (HORIZON-CL4-2022-QUANTUM02-SGA, Grant Agreement 101113690). Pasqal acknowledges the usage of the Ruby machine, the installation of which was supported by the European High-Performance Computing Joint Undertaking (JU) under grant agreement No 101018180 and project name HPCQS. The HPCQS project has received funding from the European High-Performance Computing Joint Undertaking (JU) under grant agreement No 101018180. The JU receives support from the European Union’s Horizon 2020 research and innovation programme and Germany, France, Italy, Ireland, Austria and Spain in equal parts.

\appendix

\section{Two-photon excitation scheme}\label{app:2level}
The level transition to Rydberg states involves three atomic levels $\ket{g}$, $\ket{i}$, and $\ket{r}$, coupled by two lasers with Rabi frequencies $\Omega_1$ and $\Omega_2$, respectively, and detuned by $\delta_1$ and $\delta_2$. The Rabi frequency is related to the laser intensity by $\Omega \propto \sqrt{I}$, while the detuning $\delta = \omega - \omega_0$ refers to the frequency difference between the laser frequency $\omega$ and a given atomic transition frequency $\omega_0$. To dynamically control the drive parameters, pulse-shaping devices capable of manipulating the lasers' intensity and frequency are used. 
Under the condition $\delta_1 \gg \Omega_1(t),\Omega_2(t)$, this scheme can be approximated as a two-level system described by an effective control Hamiltonian for each individual atom $i$
\begin{equation}
    \hat H_{c}^{(i)}(t) = \frac{\hbar\Omega(t)}{2}\left( \cos{\phi(t)}\hat{\sigma}^i_x-\sin{\phi(t)}\hat{\sigma}^i_y\right)-\delta(t) \hat n_i,
    \label{eq:control_Hamiltonian}
\end{equation}
where $\hat{\sigma}^i_\mu$ ($\mu\in(x,y,z)$) are the standard Pauli matrices for atom $i$, and $\hat{n}_i=(1+\hat{\sigma}^i_z)/2$. The parameters $\Omega(t)$ and $\delta(t)$ represent, respectively, the global effective Rabi frequency and detuning~\cite{de_leseleuc_analysis_2018}. The resulting effective two-level description leads to the following relations between the physical laser parameters and the effective ones:
\begin{equation}
\begin{split}
    \Omega(t) & =\frac{\Omega_1(t)\Omega_2(t)}{2\delta_1}\\ 
    \delta(t) & =\delta_2(t)+\delta_{LS}(t)\\ \phi(t) & =\phi_1(t)-\phi_2(t)
    \end{split}
    \label{eq:3to2}
\end{equation}
where $\delta_{LS}(t) = \frac{\Omega_1(t)^2-\Omega_2(t)^2}{4\delta_1}$ is the induced AC Stark shift, or light shift. While this term makes the detuning dependent on the lasers intensities, in practice our device is carefully calibrated in order to include the contribution of this term in the effective detuning. 
From this approximation, one can build the full Hamiltonian for a system of $N$ atoms as in the main text (Eq.~\eqref{eq:rydberg_ham}).

\section{Rydberg Hamiltonian and Ising model}\label{app:ising}
Starting from Eq.~\eqref{eq:rydberg_ham} and using
$\hat{n}_i = \frac{1+\hat{\sigma}_i^z}{2},$
one obtains
\begin{equation}
\hat{H}(t)
=\sum_i \frac{\hbar\Omega(t)}{2}\hat{\sigma}_i^x
+\sum_{i<j}\frac{U_{ij}}{4}\hat{\sigma}_i^z\hat{\sigma}_j^z-\sum_i h_i(t)\hat{\sigma}_i^z
+E_0(t),
\end{equation}
with
\begin{equation}
\begin{aligned}
h_i(t)&= \frac{\hbar\delta(t)}{2}-\frac{1}{4}\sum_{j(\neq i)}U_{ij},\\
E_0(t)&= -\frac{N\hbar\delta(t)}{2}+\frac{1}{4}\sum_{i<j}U_{ij}.
\end{aligned}
\end{equation}
This is a transverse-field Ising model with long-range couplings $J_{ij}=U_{ij}/4$ and transverse field $\Gamma(t)=\hbar\Omega(t)/2$. The mapping also generates longitudinal local fields $h_i(t)$, coming from both the detuning and the single-spin contributions hidden in $\hat n_i\hat n_j$. These fields are uniform only in an ideal homogeneous system; in finite or inhomogeneous geometries they become site dependent and must be taken into account explicitly.

\section{Rydberg Hamiltonian and noise parameters }\label{app:parameters_ranges}
The typical tunability range of the parameters of the Rydberg Hamiltonian in Eq.~\eqref{eq:rydberg_ham} is summarized in Table~\ref{tab:parameters} with the appropriate units of measure used throughout the paper. We use these values for the numerical emulations presented in the main text. 

The noise sources are instead summarized in Table \ref{tab:noise_sources_hamiltonian}.
We define as `classical noise' both shot-to-shot and time-dependent fluctuations of parameters defining the Hamiltonian used to implement the dynamics. In neutral atoms platforms key examples include intensity fluctuations, laser phase/frequency fluctuations and finite temperature effects such as Doppler shifts and positional disorder that effectively modulate both single-particle terms in the Hamiltonian and interaction strengths. We distinguish as `static noise' the ones that are approximately constant during one run but varying between shots and `in sequence noise' the ones that fluctuate on timescales comparable with a pulse sequence.
These errors can be generally tamed by improving laser stabilization, by having tighter traps and by cooling down the device to reduce atomic motion while they accumulate with increasing sequences duration and larger arrays of atoms.

With `coupling to the environment' we capture processes that cannot be represented as purely classical parameters fluctuations. In practice, they are summarized as decay, e.g. finite lifetime of the Rydberg state, and dephasing, e.g. due to the intermediate state finite lifetime in the typical two-photon transition.  Also in this case one expects the errors to be more relevant when increasing evolution times as the effective coupling with the environment increases.

With `state preparation and measurement' (SPAM) errors we denote imperfect initialization, e.g. incomplete optical pumping to the $\ket{g}$ state, atom loss during the sequence and readout misclassification. While measurement errors do not change the underlying coherent dynamics and can be partially corrected in the case of local observables by means of classical post-processing, it can be particularly detrimental when interested in bitstring distributions, for example in the case of solutions to optimization problems. SPAM errors also tend to be more detrimental with increasing system size: as the number of sites grows, the probability that at least one site is lost or misread increases.

Finally `systematic Hamiltonian biases' encompass coherent, repeatable deviations like calibration offsets and spatial inhomogeneities that persist across several sequences. Unlike stochastic noise, these effect produce structured and stable discrepancies with respect to the expected coherent dynamics.
Thanks to frequent calibrations these errors can be usually tamed but long experiments and large atom arrays increase the noise-sensitivity of the platform.

The modified noise levels discussed in Fig.~\ref{fig:improvement} are obtained from the ones of Tab.~\ref{tab:noise_sources_hamiltonian} with the scale transformation:
\begin{equation}
 (\alpha\sigma_\Omega, \alpha\sigma_\delta, \alpha\delta_\textrm{HF}(t), \alpha^{-1}\omega_\Omega, \alpha T, \alpha\gamma_1, \alpha\gamma_2, \alpha\eta).
\end{equation}
The "Increased" noise level is obtained by setting $\alpha=2$. Conversely, the "Reduced 1" and "Reduced 2" noise levels are obtained by setting $\alpha=1/2, 1/4$, respectively.

In the next sections, we thoroughly link the noise sources listed in Tab.~\ref{tab:noise_sources_hamiltonian} to the physical mechanism at their origin.

\begin{table*}
    \centering
\tcbox[left=0mm,right=0mm,top=0mm,bottom=0mm,boxsep=0mm,
toptitle=0.5mm,bottomtitle=0.5mm,colframe=metalblue]{
    \begin{tblr}{
        colspec = {l|[metalblue]l|[metalblue]l||l|[metalblue]l|[metalblue]l},
        column{2}={brightgreen},
        column{5}={brightgreen}
    }
        \textbf{Parameter} & \textbf{units} & \textbf{Value} & \textbf{Parameter} & \textbf{units} & \textbf{Value}\\
        \hline[metalblue]
        $C_6/\hbar$ & $\text{rad}\cdot \mu\text{s}^{-1}\cdot \mu \text{m}^6$ & 865723.02 & $r_{ij}$ & $\mu$m &  $5 \leq r_{ij} < 100$ \\
        $U \equiv \frac{1}{\hbar} \frac{C_6}{r_\text{ij}^6 }$  & $\text{rad}\cdot \mu\text{s}^{-1}$ & $0 < U \leq 2\pi\cdot 8.8$ & $t$ & $\mu$s & $t \leq 6$\\
        $\Omega$ & $\text{rad}\cdot \mu\text{s}^{-1}$ & $0 \leq \Omega \leq 2\pi\cdot 2$ & $N$ & -& $\lesssim 256$\\ 
        $\delta$ &$\text{rad}\cdot \mu\text{s}^{-1}$ & $-2\pi\cdot 10 \leq \delta \leq 2\pi\cdot 20$ \\
    \end{tblr}}
    \caption{Example of parameters ranges for a Pasqal QPU. Here we consider the Rydberg level $n=60$.}
    \label{tab:parameters}
    \centering
\tcbox[left=0mm,right=0mm,top=0mm,bottom=0mm,boxsep=0mm,
toptitle=0.5mm,bottomtitle=0.5mm,colframe=metalblue]{
    \begin{tblr}{
        colspec = {l|[metalblue]l},
        row{2}={darkgreen},
        row{3}={okabebluebg},
        row{4-7}={okabebluebg},
        row{8-15}={okabeorangebg},
        row{16}={darkgreen},
        row{17-19}={okabegreenbg},
        row{20}={darkgreen},
        row{21-24}={okabepurplebg},
        row{25}={darkgreen},
        row{26-28}={white},
    }
        \textbf{Noise source} & \textbf{Explanation}  \\
        \hline[metalblue]
        \SetCell[c=2]{c}{\textcolor{white}{Classical noise in the Hamiltonian parameters}}\\
        \SetCell[c=2]{c}{\textcolor{black}{Laser}}\\
        \hline[metalblue]
        $\sigma_\Omega/\Omega\approx 0.3\%$ &  Intensity fluctuations (shot-to-shot)\\
        $ \sigma_\delta/(2\pi)\approx 8 \text{kHz}$ & Phase and intensity fluctuations (shot-to-shot)\\
        $\delta_{\text{HF}}(t)=\sum_i \sqrt{2(f_{i+1}-f_{i})S_\nu(f_i)}\cos(2\pi f_it+\varphi_i)$ & Phase fluctuations (in-sequence)\\
        $w_\Omega\approx 141\mu\text{m}$ & Spatial profile (static).  \\
        \hline[metalblue]
        \SetCell[c=2]{c}{\textcolor{black}{Temperature}}\\
        \hline[metalblue]
        $T\approx 20\,\mu\text{K}$ & Atomic trap temperature \\
        $w_\text{trap}\approx 0.84\mu\textrm{m}$ & Trap waist \\
        $U_\text{trap}\approx 70\mu\textrm{K}$ & Trap depth \\
        $\sigma^T_{\delta_i}/(2\pi)=\lVert \vec{k}_{\mathrm{eff}} \rVert\sqrt{\frac{k_B T}{m}}\approx 50 \text{kHz}$ & Thermal Doppler shifts (shot-to-shot) \\
        $\sigma^T_{r_i^{xy}}=\sqrt{\frac{k_B T w_\textrm{trap}^2}{4U_\textrm{trap}}}\approx 0.22\mu\text{m}$ & In-plane thermal disorder (shot-to-shot)\\
        $\sigma^T_{r_i^z}=\frac{\pi}{\lambda}\sqrt{2}w_\textrm{trap}\sigma^T_{r_i^{xy}}\approx 1\mu\text{m}$ ($\lambda\approx 0.85\mu\textrm{m}$)& Off-plane thermal disorder (shot-to-shot)\\
        $\sigma_{v^{xyz}}=\sqrt{\frac{k_B T}{m}}\approx 0.025\frac{\mu\text{m}}{\mu\text{s}}$ & Thermal velocity (in-sequence)$^{*}$  \\

        \SetCell[c=2]{c}{\textcolor{white}{Effective channels beyond the two-level (qubit) model}}\\
        $\gamma_{1}\approx 0.01 \mu\text{s}^{-1}$ & $T_1$ decay channel \\
        $\gamma_{2}\approx 0.05 \mu\text{s}^{-1}$ & $T_2$ dephasing channel \\
        $\{\gamma_\textrm{n}\}$ & Irreversible decays outside the $\ket{g},\,\ket{r}$ qubit manifold.$^{*}$  \\
        \SetCell[c=2]{c}{\textcolor{white}{SPAM errors}}\\
        $\eta\approx 1.8\%$ & Atom not pumped to the $\ket{g}$ state.  \\

    $\epsilon \approx 1\%$  & Misidentification of $\ket{g}$ as $\ket{r}$ \\
    $\epsilon '\approx 7\%$ & Misidentification of $\ket{r}$ as $\ket{g}$ \\
  $\epsilon_\text{shot}$ & Finite number of bitstrings \\
        \SetCell[c=2]{c}{\textcolor{white}{Calibration-limited Hamiltonian errors}}\\
        $\Delta \Omega/\Omega\approx 1\%$ & Precision of $\Omega$-calibration and slow temporal drift. \\ 
        $\Delta \delta/(2\pi) \approx 100$ kHz & Precision of $\delta$-calibration and slow temporal drift.  \\   
         $\Delta r /r\approx 1\% $ & (de)magnification of the SLM pattern 
        \end{tblr}}
    \caption{Noise sources of a neutral atom Quantum Processor Unit (QPU). For each source we provide, when available, their relation with quantities measurable in the neutral atom device and that can be input in the noise model emulator. Typical values are obtained from the QPU Pasqal FC1 during the data collection of this work. For the classical noise in Hamiltonian parameters we write explicitly its variation time scale (in-sequence, shot-to-shot, or static). $^{*}$These noise sources are not included in our numerical emulation.}
    \label{tab:noise_sources_hamiltonian}
\end{table*}

\section{Noise description}\label{app:noise}
\subsection{Thermal noise}\label{app:thermalnoise}
In this section, we focus on the impact of finite temperature on the system dynamics and on its numerical modeling. Although the atoms are confined within optical tweezers, they are not perfectly stationary~\cite{de_leseleuc_analysis_2018,tuchendler_energy_2008}. 
Due to their finite temperature $T$, the atoms retain a residual thermal motion. Along a given axis, this characterized by a velocity distribution with a standard deviation
\begin{equation}\label{eq:thermalvelocity}
\sigma_v(T) = \sqrt{\frac{k_B T}{m}},
\end{equation}
where $k_B = 1.38 \times 10^{-23}~\mathrm{J/K}$ is the Boltzmann constant and 
$m = 1.45 \times 10^{-25}~\mathrm{kg}$ is the atomic mass ($^{87}$Rb in this work). This residual motion introduces shot-to-shot fluctuations in the effective Hamiltonian parameters, through the Doppler shift and positional disorder.\footnote{At low temperatures, such that $k_{B}T\ll\hbar \omega$, the classical approach breaks down and velocity standard deviation becomes 
\begin{equation}
    \sigma_{v} = \sqrt{\frac{\hbar\omega}{2m}\coth\left(\frac{\hbar\omega}{2k_{B}T}\right)}.
    \end{equation}
The position distribution is then given by $\sigma_r = \frac{\sigma_{v}}{\omega}$.}\\

\paragraph{Doppler noise -} Each atom $i$ has a thermal velocity $\vec{v}_i$ drawn independently from a Gaussian distribution with standard deviation $\sigma_v(T)$ given in Eq.~\eqref{eq:thermalvelocity}. As a result, each atom experiences a different effective laser frequency due to the Doppler effect, introducing shot-to-shot and atom-dependent detuning fluctuations $\delta \to \delta_i$. The standard deviation of the resulting detuning spread is:
\begin{equation}
\sigma_\delta^{T} = \lVert \vec{k}_{\mathrm{eff}} \rVert \, \sigma_v(T),
\end{equation}
where $\vec{k}_{\mathrm{eff}} = 2\pi \times 1.4~\mathrm{\mu m}^{-1}$ is the combined wave vector of the counter-propagating laser system. To assess the impact of these Doppler-induced fluctuations, the magnitude of the detuning spread $\sigma_\delta^{T}$ must be compared to the Rabi frequency $\Omega$. If $\sigma_\delta^{T} \ll \Omega$, the effect is negligible. Otherwise, we sample each detuning $\delta_i$ at every shot from a Gaussian distribution centered on $\delta$, with standard deviation $\sigma_\delta^{T}$. \\

\paragraph{Positional disorder -} Thermal motion also contributes to dephasing between experimental shots by inducing shot-to-shot variations in the atomic positions. The associated positional spread can be related to the velocity distribution of Eq.~\eqref{eq:thermalvelocity} through
\begin{equation}
\sigma_r({T}) = \frac{\sigma_v(T)}{\omega_{\textrm{rad}}},
\end{equation}
where $\omega_{\textrm{rad}} \approx 2\pi \times 0.1~\mathrm{MHz}$ is the typical radial angular trapping frequency of the tweezers.

This residual motion modifies the system Hamiltonian through two main effects: fluctuations of the interaction strength among Rydberg atoms, and spatial dependence of the intensity profile of the lasers.
The interaction between atoms depends on their mutual separation through the relation $U(r_{ij}) = C_6/r_{ij}^6$.  Even for small positional spread $\sigma_r^T$ compared to the interatomic distance $r_{ij}$, thermal motion leads to non-negligible shot-to-shot fluctuations of the interaction strength. Specifically, we get $\frac{\sigma_U}{U(r_{ij})} \approx 6 \frac{\sigma_{r_{ij}}}{r_{ij}}$ to first order. 

This effect can be simulated by drawing a position, for every atom and at every shot, from the correct thermal distribution. 
In our noise model, the in-plane ($x$–$y$) and out-of-plane ($z$) position fluctuations are sampled from normal distributions centered at the nominal trap positions, with standard deviations $\sigma^T_{r_i^{xy}}$ and $\sigma^T_{r_i^{z}}$, respectively.
Both can be connected to the temperature and trap parameters as
\begin{equation}
    \sigma^T_{r_i^{xy}}=\sqrt{\frac{k_B T w_\textrm{trap}^2}{4U_\textrm{trap}}},\ \ \ 
    \sigma^T_{r_i^{z}}=\frac{\pi}{\lambda}\sqrt{2}w_\textrm{trap}\sigma^T_{r_i^{xy}}.
    \end{equation}
Here $U_\textrm{trap}$ is the trap depth, $w_\textrm{trap}$ is the trapping beam waist, $\lambda$ is the trapping wavelength, and $T$ is the atomic temperature. 
The equations above can be derived by treating the atom in the optical tweezer as a classical harmonic trap at thermal equilibrium~\cite{grimm_optical_2000}. Note that for global Rydberg beams, the  laser waists are far larger than the scale of this distribution (\(\frac{\omega_{\Omega, \delta}}{\sigma_r} \approx 10^3\)), and the laser parameters can be considered constant for each atom.\\
\subsection{Controls}\label{app:controls}
Real lasers are not perfectly stable and exhibit both intensity and phase fluctuations \cite{stephan_laser_2005,domenico_simple_2010}, which can significantly affect driven atomic dynamics. These fluctuations can be separated into slow (low-frequency) and fast (high-frequency) contributions. High-frequency noise occurs on timescales comparable to or shorter than the pulse duration and is described by the relative intensity- or frequency-noise power spectral density (PSD)~\cite{jiang_sensitivity_2023,clade_oscillations_2005}, which can be inferred from experimental measurements.

The low-frequency part of the PSD is often difficult to characterize experimentally. We therefore truncate the PSD below a cutoff frequency and treat the corresponding fluctuations as shot-to-shot variations. Fluctuations with frequencies satisfying $1/f \gg t$, where $t$ is the processing time, can be regarded as constant during a given experimental realization. \\

\paragraph{Laser intensity fluctuations -} Fluctuations in the laser intensity $I$ directly translate into fluctuations of the Rabi frequency, since $\Omega \propto \sqrt{I}$. This results in errors in the pulse area, dephasing of Rabi oscillations, and more generally decoherence \cite{de_leseleuc_analysis_2018}. In multi-level excitation schemes, intensity noise also modulates AC Stark shifts, producing effective detuning noise.
In numerical simulations, the high-frequency intensity noise is incorporated by generating a stochastic relative correction $I \rightarrow I\cdot(1+\alpha_{HF} (t))$ where
\begin{equation}
    \alpha_{\rm HF}(t) = \sum_i \sqrt{2(f_{i+1}-f_i) S_I(f_i)} \cos(2\pi \cdot f_i \cdot t + \varphi_i).
\end{equation}
Here, $S_I$ is the intensity noise PSD and $\varphi_i$ are random phases sampled from $[0,2\pi)$ for each discrete frequency point $f_i$~\cite{clade_oscillations_2005}.
The low-frequency intensity fluctuations are modeled by a shot-to-shot random variation of the Rabi frequency, characterized by a standard deviation $\sigma_{\Omega}^{\rm shot}$.  This deviation is typically of order of a few percent (see Tab.~\ref{tab:noise_sources_hamiltonian}).\\

\paragraph{Laser frequency fluctuations -} While narrow, the spectrum of a laser is not infinitely narrow, and a laser naturally has some excursions around its carrier frequency \cite{stephan_laser_2005,domenico_simple_2010}. This can be characterized either by phase or frequency noise and leads to a gradual loss of phase coherence between the driving field and the atomic system, resulting in decoherence and effective broadening of the transition~\cite{day_limits_2022}.  
Again, phase fluctuations are characterized by the frequency-noise power spectral density (PSD), that we denote as $S_\nu(f)$~\cite{clade_oscillations_2005,jiang_sensitivity_2023}.
The high-frequency noise introduces a correction term in the detuning setpoint of the pulse following
\begin{equation}
    \delta_{\rm HF}(t) = \sum_i \sqrt{2(f_{i+1}-f_i) S_\nu(f_i)} \cos(2\pi f_i t + \varphi_i),
\end{equation}
where $\varphi_i \in  [0,2\pi)$. Note that the sum here is assumed to run over positive frequency components, and that we assume a single-sided PSD. Using a two-sided PSD, would introduce an additional factor $\sqrt{2}$. 
For completeness, let us note that one can talk about frequency and phase noise interchangeably, since the PSDs are simply related through the relation $S_\nu(f) = f^2S_\phi(f)$.\\

\subsection{Effective couplings}\label{app:channels}
Like any many-body quantum system, atoms are also inherently coupled to environmental degrees of freedom not explicitly included in the system Hamiltonian. This coupling leads to decoherence, which limits the fidelity and coherence time of quantum operations. Here we list several physical mechanisms that contribute to this effect.
Rydberg states have a finite lifetime due to spontaneous emission and coupling to the environment's black-body radiation (BBR). At an environmental temperature of zero, the lifetime scales approximately as $\tau(n;T=0\,\mathrm{K}) \propto n^3$, where $n$ is the principal quantum number. The associated decoherence rate is
\begin{equation}
\Gamma_{r\rightarrow g} = 1/\tau(n;T=0\,\mathrm{K}),\quad \mathrm{e.g.},\quad \tau(n=60)\approx 230.3~\mu\mathrm{s}.
\end{equation}
At finite environmental temperatures, BBR induces transitions between nearby Rydberg states, reducing the lifetime. This thermal contribution is described by 
\begin{equation}
\Gamma_{r\rightarrow r^\prime}(T) = \frac{1}{\tau(n;T)} - \frac{1}{\tau( n;T=0\,\mathrm{K})},
\end{equation} 
where, e.g.,  $\tau(n=60;T=300\,\mathrm{K}) \approx 100.6~\mu\mathrm{s}$.
The typical two-photon excitation from the ground state $\ket{g}$ to the Rydberg state $\ket{r}$ involves an intermediate state $\ket{i}$, which, despite being far detuned, gets admixed into the others with populations $\Omega_1^2/4\delta_1^2$ and $\Omega_2^2/4\delta_1^2$ for the $|g\rangle$ and $|r\rangle$ states, respectively. Due to the intermediate state's short lifetime ($1/\gamma_i\sim 120~\mathrm{ns}$ for Rb's $6P_{3/2}$ state), even relatively small admixtures can contribute significantly to the observed dephasing. This decoherence channel can be modeled through jump operators with prefactors equaling $\sqrt{\gamma_i\Omega_{1,2}^2/4\delta_1^2}$, although the exact form of the jump operator depends on the magnetic sublevels that are chosen, and the decay paths that are allowed by selection rules.

The combination of decay and dephasing makes the dynamics of the system non-unitary. Formally, this can be described by means of a Lindblad master equation for the density matrix $\rho$:
\begin{equation}
\dot \rho = -i[\hat H, \rho] + \sum_k \gamma_k \left( \hat L_k \rho \hat L_k^\dag - \frac{1}{2} \{ \hat L_k^\dag \hat L_k, \rho \} \right),
\end{equation}
where $[\cdot,\cdot]$ and $\{\cdot,\cdot\}$ denote the commutator and anti-commutator, respectively.
Furthermore, $\hat L_k$ are the jump operators associated with each dissipative channel and $\gamma_k$ are the corresponding rates. The relevant channels described before can be exemplified like follows: $(i)$ dephasing is represented by the operator $\hat L = \ket{r}\bra{r}$ with rate $\gamma = \Gamma_\mathrm{deph}^{\mathrm{eff}}$, $(ii)$ decay mechanisms from state $\ket{a}$ to state $\ket{b}$ can be described by $\hat L = \ket{b}\bra{a}$ with rate $\gamma = \Gamma_{a\rightarrow b}$.

In summary, coupling of Rydberg atoms to their environment gives rise to effective decay and dephasing channels that can be incorporated into a Lindblad formalism, enabling efficient simulation of the open-system dynamics.
In practice, the Lindblad master equation can be unraveled using quantum trajectories~\cite{daley_quantum_2014,breuer_concepts_2003,fazio_many-body_2025},
which propagate pure states under a non-Hermitian effective Hamiltonian interrupted by stochastic quantum jumps.

In addition to the decay channels above, refined noise models can include leakage channels, in which population escapes the computational basis but is still read out as a computation state. For example, an atom in $\ket{r}$ might decay into  a long-lived Rydberg state $\ket{r'}$ that is uncoupled from the drive and thus detected as $\ket{r}$. Likewise, scattering from the intermediate level can transfer population to a dark ground state $\ket{g'}$, that does not take part in the dynamics and is detected as $\ket{g}$. These processes can be represented by additional jump operators into auxiliary states, together with the corresponding branching ratios and detection mappings.

\subsection{SPAM errors}\label{app:SPAM}
\paragraph{State preparation errors -} They occur when an atom is initially pumped into an unwanted hyperfine ground state that is not coupled to the Rydberg state by the two-photon transition. We denote the probability of this occurring per site and per shot as $\eta$. In simulations, this can be implemented by randomly selecting sites that are neglected in the many-spin dynamics with a probability $\eta$. For these sites, the final bitstring always corresponds to the $\ket{g}$ state. A more refined treatment could be including detuning effects to account for the dynamics of atoms initialized in different hyperfine states, though this is often a small correction.\\

\paragraph{Measurement errors -} At the measurement stage two different errors might happen. We model these errors as independent classical misidentification processes acting on each atom. A false positive error (false Rydberg detection in Tab.~\ref{tab:noise_sources_hamiltonian}
), occurring with probability $\varepsilon$, corresponds to an atom that is actually in the state $\ket{g}$ being detected as $\ket{r}$. In experiments, this error predominantly originates from atom loss during the experimental sequence or during fluorescence imaging, for instance due to collisions with residual background gas or thermal motion. Conversely, a false negative error (false ground-state detection in Tab.~\ref{tab:noise_sources_hamiltonian}), occurring with probability $\varepsilon'$, corresponds to an atom that is actually in the state $\ket{r}$ being detected as $\ket{g}$. This error typically arises when an atom excited to the Rydberg state decays back to the ground state prior to detection and is subsequently recaptured. 
These errors can be incorporated in simulations by randomly flipping the measured bitstrings according to the corresponding probabilities.
While these measurement errors can be very harmful when the goal is to obtain a particular bitstring, the impact of these errors in the expectation value of a single or two-body correlator can be successfully mitigated if we know their value from previous calibrations.
Due to errors, the measured single-site occupation $\widetilde{\langle n_i\rangle}$ can be written as
\begin{equation}
\widetilde{\langle n_i\rangle} = (1-\varepsilon')P_{r} + \varepsilon P_g
= (1-\varepsilon-\varepsilon')\langle n_i\rangle + \varepsilon,
\end{equation}
where $P_{g}$($P_{r}$) are the exact probabilities of measuring an atom in the ground state $\ket{g}$ (rydberg state $\ket{r}$), respectively.
The occupation $\widetilde{\langle n_i\rangle}$ can be inverted to obtain the corrected expectation value
\begin{equation}
\label{eq:spam_single}
\langle n_i\rangle = P_{r}=\frac{\widetilde{\langle n_i\rangle}-\varepsilon}{1-\varepsilon-\varepsilon'}.
\end{equation}

For two-point quantities such as $\langle n_i n_j\rangle$ we can define a measurement error matrix
\begin{equation}
\begin{bmatrix} \tilde{P}_{rr} \\ \tilde{P}_{gg} \\ \tilde{P}_{gr/rg} \end{bmatrix}
 = \mathbf{M}
\begin{bmatrix} P_{rr} \\ P_{gg} \\ P_{gr/rg} \end{bmatrix},
\end{equation}
where the transition matrix $\mathbf{M}$ is defined as:
\begin{equation}
\mathbf{M} = 
\begin{bmatrix}
(1-\varepsilon')^2 & {\varepsilon}^{2} & (1-\varepsilon')\varepsilon \\
\varepsilon'^{2} & (1-\varepsilon)^{2} & \varepsilon'(1-\varepsilon) \\
2\varepsilon'(1-\varepsilon') & 2\varepsilon(1-\varepsilon) &
1-\varepsilon'-\varepsilon + 2\varepsilon'\varepsilon
\end{bmatrix}.
\end{equation}
By inverting this matrix, we obtain the corrected probabilities, from which the two-body occupation correlator follows directly as
\begin{equation}
\langle n_i n_j\rangle = P_{rr}=\frac{(1-\varepsilon)^2\,\tilde P_{rr}+\varepsilon^2\,\tilde P_{gg}-\varepsilon(1-\varepsilon)\,\tilde P_{gr/rg}}
{(1-\varepsilon-\varepsilon')^2}.
\end{equation}

In some applications, most notably combinatorial optimization, one is less interested in estimating expectation values and more in the sampled bitstring distribution itself, or in obtaining bitstrings that minimize a given cost function. In this setting it is common to apply a purely classical postprocessing step to the raw measurement outcomes. The approach consists in performing rounds of local bit flips together with validity checks to reduce the associated cost of the bitstrings, guaranteeing the feasibility of the solution. Such heuristic procedures are routinely used for the maximum weighted independent set (MWIS) problem and more general quadratic unconstrained binary optimization (QUBO) instances~\cite{cazals_identifying_2025,ebadi_quantum_2022,leclerc_implementing_2025}.

\subsection{Systematic off-sets and biases}
Systematic errors arise from our inability to perfectly control the Hamiltonian parameters in both space and time. Physically, they correspond, for instance, to (constant) spatial inhomogeneities or deviations in the time-dependence of the effective Hamiltonian parameters $\Omega$, $\delta$, and the interaction strength. They remain constant across shots and persist until the setup is recalibrated or the hardware is upgraded (e.g., by improving the precision with which the laser power is controlled). These errors do not lead to decoherence through shot-to-shot averaging, but instead change the implemented dynamics by introducing constant biases. They can therefore be handled during the algorithm design stage by incorporating these effects into the model or by assessing robustness to shifts in the control parameters, without requiring Monte-Carlo sampling.\\

\paragraph{Laser miscalibration -}
A miscalibration, or a drift of the laser system between calibration and the actual experimental sequence, results in systematic biases of the drive parameters, such that $\Omega \rightarrow \Omega + \Delta\Omega$ and $\delta \rightarrow \delta + \Delta\delta$ in the effective Ising Hamiltonian (Eq.~\ref{eq:rydberg_ham}). These biases shift the operating point of the system, leading to systematic errors in pulse areas and effective detuning. Typical offsets for $\Omega$ and $\delta$ are reported in Table~\ref{tab:noise_sources_hamiltonian} and are usually $\Delta \Omega/\Omega \sim 1\%$ and $\Delta \delta/(2\pi) \sim 100~\mathrm{kHz}$.\\

\paragraph{Laser-beam spatial profile -} In our setup, the excitation beams have a Gaussian spatial profile, leading to spatial inhomogeneities of the control parameters across the atomic array. The beams propagate along $y$, while the atomic register lies in the $(x,y)$ plane, with a typical extent of $\sim 100~\mu$m. For a circular Gaussian beam of waist $w_0$, the electric-field amplitude can be written as:
\begin{equation}
    \mathcal{E}\propto\frac{w_0}{w(y)}\exp \left(-\frac{x^2+z^2}{w(y)^2}\right),
\end{equation}
with $w(y)=w_0\sqrt{1+\frac{y^2}{y_r^2}}$, and $y_r=\pi w_0^2/\lambda$ the Rayleigh range ($\lambda$ being the wavelength of the light)~\cite{saleh_fundamentals_1991}. Since $y_r$ is typically on the cm scale for our beam parameters, variations along the propagation axis $y$ are negligible over the array size, and the dominant inhomogeneity arises from the transverse profile in $(x,z)$. In the focal plane ($y=0$), the effective Rabi frequency for aligned excitation beams remains Gaussian:
\begin{equation}
\Omega\propto \Omega_1 \Omega_2 \propto \exp\left(-\frac{x^2+z^2}{w_{\Omega}^2}\right),
\end{equation}
where $w_{\Omega}=w_1w_2/\sqrt{w_1^2+w_2^2}$ is the effective waist (Cf. Table~\ref{tab:noise_sources_hamiltonian}) obtained from the individual beam waists $w_1$ and $w_2$.\\

\paragraph{SLM pattern errors -} The atom trap layout is generated using a Spatial Light Modulator (SLM). Due to its finite resolution, traps cannot be positioned with arbitrary precision~\cite{catala-castro_positioning_2021,engstrom_calibration_2013,barredo_synthetic_2018,labuhn_tunable_2016,curtis_dynamic_2002}. This introduces a systematic error in the position of each trap site, which is unchanged until the layout is re-calibrated. This error is small compared to the thermal spread in the atoms' position for the typical temperatures given earlier. 

In addition, uncertainties in the imaging magnification used to calibrate the layout introduce a systematic error on interatomic distances. Overall, these effects lead to a static relative error on interatomic spacings, typically $\Delta r/r \sim 1\%$ (see Table~\ref{tab:noise_sources_hamiltonian}).

These errors on trap positions and interatomic distances directly impact the van der Waals interaction, and becomes especially critical when the interaction strength is comparable to the Rabi frequency. \\

\paragraph{Finite control bandwidth -}
Temporal control of the laser pulses that we use is achieved through devices that can manipulate light on a fast enough time scale. Here we discuss these devices and explain the limitations they impose on the implemented sequences. One example is an acousto-optic modulator, that uses the acousto-optic effect in which an acoustic wave modulates the index of refraction of a crystal through which the light propagates \cite{maydan_acoustooptical_1970}. The laser beam diffracts off of this, and the amplitude and frequency of the running wave respectively control the laser's Rabi frequency and detuning.

The group velocity of the acoustic wave is finite, resulting in a low-passing behavior as far as the modulation is concerned. The response timescale is set by the time it takes for the acoustic wave to propagate through the beam spot. More rigorous expressions for the bandwidth can be derived as well, see, e.g. Ref.~\cite{maydan_acoustooptical_1970}. Typical corner frequencies are on the order of 5 MHz, corresponding to response times of 200 ns.

Faster control can be achieved using electro-optic modulators which have a birefringence that depends on an applied electric field \cite{hecht_optics_2017}. When combined with additional polarization optics, they can be turned into devices that dynamically control the laser amplitude -- and with that the Rabi frequency. EOMs allow for relatively fast control, with bandwidths reaching up to 100 MHz (corresponding to 10 ns rise times). A downside is that they cannot readily be used to control the laser frequency.

In multi-level excitation schemes, one thing to keep in mind is that the net control bandwidths of the various laser channels are not necessarily matched, which makes it difficult to precisely compensate the light shifts of Eq.~(\ref{eq:3to2}) in the presence of rapidly changing parameters. Sometimes this can result in detuning transients that, when integrated over time, result in effective phase differences between the drive and the atoms. In particular, when one of the lasers is switched quickly using an EOM, the light shift this results in through Eq.~(\ref{eq:3to2}) can only be matched at one set of parameter values, the reason being that frequency control is typically slower since it's implemented using AOMs.

\section{Numerical emulation}\label{app:numerics}
In this work we implement the noise model described in Sec.~\ref{sec:qpu_noise} using the \texttt{emu-mps} library~\cite{bidzhiev_efficient_2025}, a matrix-product-state backend to emulate QPU pulse sequences generated in the \texttt{Pulser} library~\cite{silverio_pulser_2022}. The same backend is also used for the ideal (noiseless) emulations. To investigate the dynamics of $6\times 6$ qubit arrays we use a small $dt$ as a time step in the time-dependent variational principle used by the emulator, and a maximum bond dimension of $\chi\leq 512$. This choice of bond dimension represents a truncation of the Hilbert space, which is already intractable with raw state-vector methods at this system. That is, the full dimension is $2^{36}\simeq 6.9 \times 10^{10}$, which would require a bond dimension $\chi_\textrm{exact}\approx 2^{18}$. Nevertheless, we use finite-bond dimension scaling to check that for such early-time dynamics the impact of this truncation is in practice negligible for the protocols under consideration. This highlights the versatility of tensor-network methods to provide noiseless and noisy benchmarks of the QPU performance at system sizes lying at the edge of classical simulability. Noise effects are generally estimated by averaging over 100 Monte Carlo trajectories. Uncertainty bands for the noise-model predictions are constructed from the 2.5th–97.5th percentile range of simulated trajectories, corresponding approximately to a two-standard-deviation interval for Gaussian fluctuations. Below, we discuss specific methodologies for the annealing and post-quench dynamics.
\subsection{Annealing protocols}
\begin{figure}
    \centering
    \includegraphics[width=0.8\linewidth]{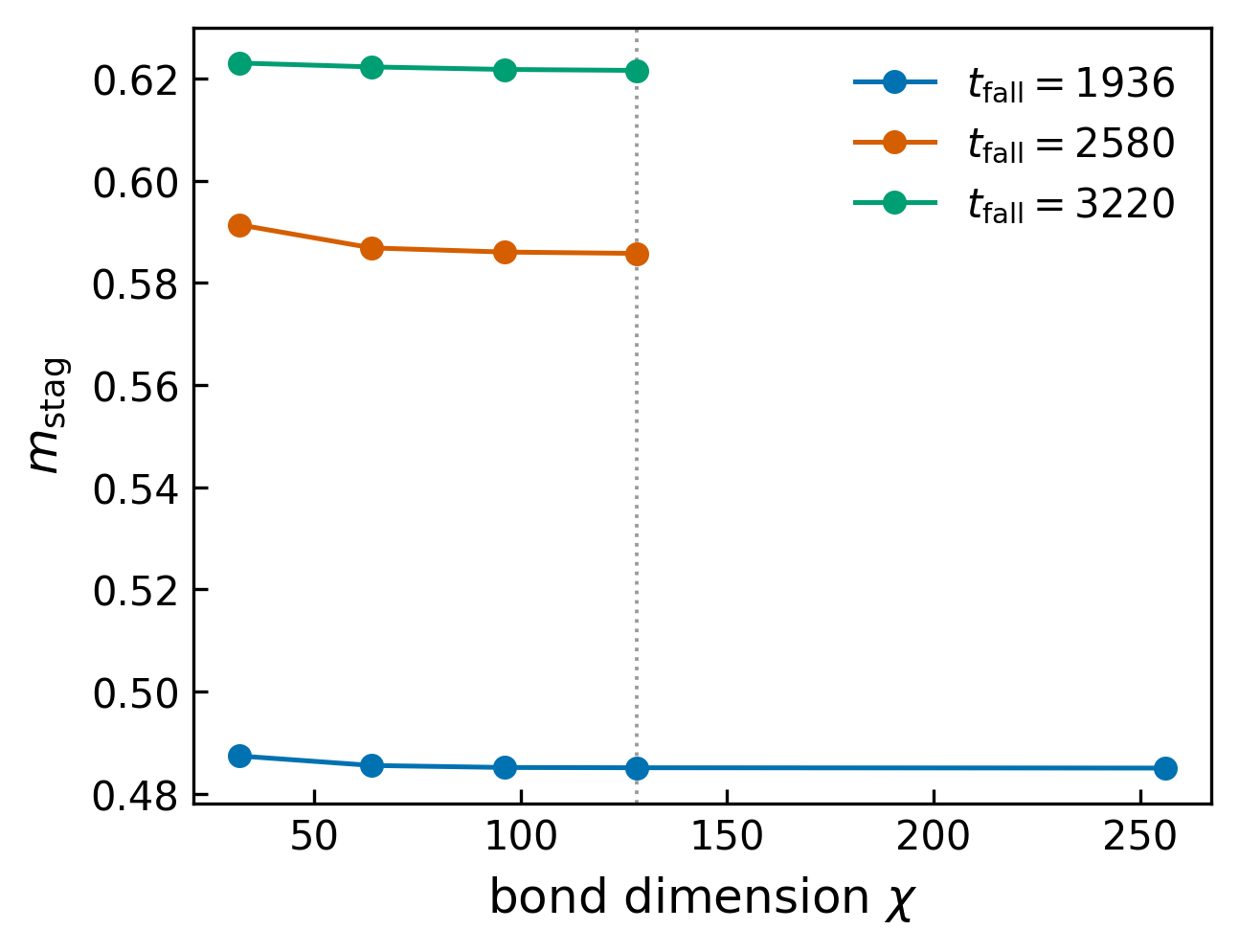}
    \caption{Noiseless MPS-simulated staggered magnetization $m_\mathrm{stag}$ as a function of the maximum bond dimension $\chi$, for the $6\times6$ annealing protocol at three representative ramp durations $t_\mathrm{fall} = 1936$, $2580$, and $3220,$ns (blue, orange, green). $m_\mathrm{stag}$. For each $t_\mathrm{fall}$, $m_\mathrm{stag}$ varies by less than $0.5\%$ between $\chi=128$ and $\chi=256$ (dashed vertical line marks $\chi=128$), confirming that $\chi=128$ is sufficient for convergence; this value is used throughout the main text.}
    \label{fig:scaling_annealing}
\end{figure}

For the annealing protocol, the relevant quantity is the ordered bitstring prepared at the end of the sequence rather than the transient dynamics. We therefore sample $N_\mathrm{shots}=10^3$ bitstrings from the final matrix-product state of each trajectory, reproducing the projective readout of the QPU, and estimate the staggered magnetization from their statistics; the noise model is averaged over $100$ Monte Carlo trajectories for the full model and $40$ trajectories for each isolated noise family. Since the target here is a low-entanglement ordered state rather than a highly entangled dynamical one, a smaller bond dimension of $\chi=128$ and a coarser time step $dt=10\,$ns already converge the observable, see Fig.~\ref{fig:scaling_annealing}.
\subsection{Post-quench protocols}
\begin{figure}
    \centering
    \includegraphics[width=\columnwidth]{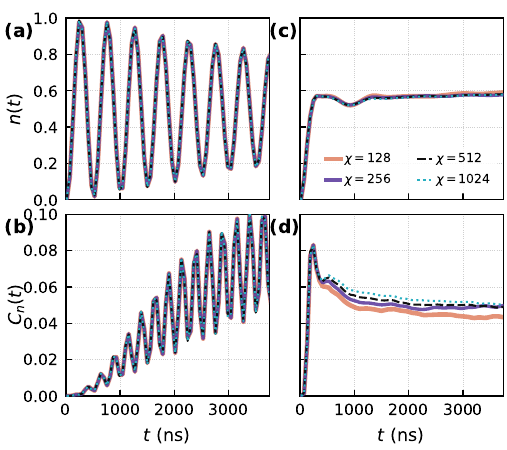}
    \caption{Noiseless behavior of observables during post-quench dynamics, computed with \texttt{emu-mps}. Different bond dimensions ($\chi=128,\,256,\,512,\,1024$) are shown for comparison. (\textbf{a},\textbf{b}) Local population and nearest-neighbor correlations in the small interaction regime, $\Omega/U=14.5$.  (\textbf{c},\textbf{d}) Local population and nearest-neighbor correlations in the strong interaction regime, $\Omega/U=1.1$. }
    \label{fig:scaling_quenches}
\end{figure}

For the post-quench protocol, the relevant quantities are expectation values of physical observables over time. For computational efficiency, we estimate the latter directly from end-time trajectory wavefunctions. That is, we avoid sampling bitstrings and computing independent trajectories ensembles at each intermediate time. Due to the higher numerical complexity of this protocol, compared to annealing dynamics, we use stricter parameters, $dt=5\,$ns and $\chi=512$. To benchmark the impact of the $\chi=512$ truncation, in Fig.~\ref{fig:scaling_quenches} we show the dependence with the truncated bond dimension of the observables considered in the main text. On the one hand, we observe a good convergence already at $\chi=512$ in the small interaction regime, both for the local population, panel (a), and connected correlations, panel (b). On the other hand, the interacting regime shows a good convergence in the local population dynamics, panel (c), while we observe a finite scaling in connected correlations, panel(d).

\end{document}